\documentclass[fleqn,usenatbib]{mn2e}
\usepackage[letterpaper, top = 0.95in, bottom=0.95in, left=0.95in, right=0.95in]{geometry}

\usepackage{amsmath}
\usepackage{txfonts}
\usepackage{graphicx}
\usepackage{threeparttable}
\usepackage{latexsym}
\usepackage{amsbsy}
\usepackage{float}
\usepackage{mathrsfs}
\usepackage{color}

\begin{document}

\title[Cosmological Constraints From Weak Lensing Peak Statistics With CFHT Stripe 82 Survey]{Cosmological Constraints From Weak Lensing Peak Statistics With CFHT Stripe 82 Survey}

\author[X. Liu et al.]
{Xiangkun Liu,$^{1}$\thanks{Email: lxk98479@pku.edu.cn}
Chuzhong Pan,$^{1}$
Ran Li,$^{2}$
Huanyuan Shan,$^{3}$ \newauthor
Qiao Wang,$^{2}$
Liping Fu,$^{4}$ 
Zuhui Fan,$^{1,5}$\thanks{Email: fanzuhui@pku.edu.cn}
Jean-Paul Kneib,$^{3,6}$ \newauthor
Alexie Leauthaud,$^{7}$
Ludovic Van Waerbeke,$^{8}$
Martin Makler,$^{9}$
Bruno Moraes,$^{10,11}$ \newauthor
Thomas Erben,$^{12}$
Ald\'ee Charbonnier$^{13}$ \\ \\
$^1$Department of Astronomy, Peking University, Beijing 100871, China,\\
$^2$Key Laboratory for Computational Astrophysics, The Partner Group of Max Planck Institute for Astrophysics, National Astronomical Observatories, \\
Chinese Academy of Sciences, Beijing, 100012, China\\
$^3$Laboratoire d'astrophysique (LASTRO), Ecole Polytechnique F\'ed\'erale de Lausanne (EPFL), Observatoire de Sauverny, CH-1290 Versoix, Switzerland\\
$^4$Shanghai Key Lab for Astrophysics, Shanghai Normal University, Shanghai 200234, China\\
$^5$Collaborative Innovation Centre of Modern Astronomy and Space Exploration, China\\
$^6$Aix Marseille Universit\'e, CNRS, LAM (Laboratoire d'Astrophysique de Marseille) UMR 7326, 13388, Marseille, France\\
$^7$Kavli Institute for the Physics and Mathematics of the Universe (Kavli IPMU, WPI), 
The University of Tokyo, Kashiwa, Chiba 277-8582, Japan\\
$^8$Department of Physics and Astronomy, University of British Columbia, 6224 Agricultural Road, Vancouver, V6T 1Z1, BC, Canada\\
$^9$Centro Brasileiro de Pesquisas F\'isicas, Rua Dr. Xavier Sigaud 150, Rio de Janeiro, RJ 22290-180, Brazil\\
$^{10}$Department of Physics and Astronomy, University College London, Gower Street, London, WC1E 6BT, UK\\
$^{11}$CAPES Foundation, Ministry of Education of Brazil,  Brasilia/DF 70040-020, Brazil\\
$^{12}$Argelander Institute for Astronomy, University of Bonn, Auf dem H{\"u}gel 71, 53121 Bonn, Germany\\
$^{13}$Observat\'orio do Valongo, Universidade Federal do Rio de Janeiro, Ladeira do Pedro Ant\^onio 43, Sa\'ude, Rio de Janeiro, RJ 20080-090, \\
Brazil \& Centro Brasileiro de Pesquisas F\'isicas, Rua Dr. Xavier Sigaud 150, Rio de Janeiro, RJ 22290-180, Brazil}

\maketitle

\begin{abstract}
{
We derived constraints on cosmological parameters using weak lensing peak statistics
measured on the $\sim130\deg^2$ of the Canada-France-Hawaii Telescope Stripe 82 Survey (CS82). This analysis demonstrates the feasibility of using peak statistics in cosmological studies. For our measurements, we considered peaks with signal-to-noise ratio in the range of $\nu=[3,6]$. For a flat $\Lambda$CDM model with only $(\Omega_{\rm m}, \sigma_8)$ as free parameters, we constrained the parameters of the following relation $\Sigma_8=\sigma_8(\Omega_{\rm m}/0.27)^{\alpha}$ to be: $\Sigma_8=0.82 \pm 0.03 $ and $\alpha=0.43\pm 0.02$.
The $\alpha$ value found is considerably smaller than the one measured in
two-point and three-point cosmic shear correlation analyses, showing a
significant complement of peak statistics to standard weak lensing cosmological studies.
The derived constraints on $(\Omega_{\rm m}, \sigma_8)$ are fully consistent with the ones from either WMAP9 or Planck.
From the weak lensing peak abundances alone, we obtained marginalised mean values of
$\Omega_{\rm m}=0.38^{+0.27}_{-0.24}$ and $\sigma_8=0.81\pm 0.26$. 
Finally, we also explored the potential of using weak lensing peak statistics to constrain the mass-concentration relation of dark matter halos 
simultaneously with cosmological parameters.
}
\end{abstract}
\begin{keywords}
cosmology - dark matter - clusters: general - gravitational lensing: weak - large-scale structure of universe.
\end{keywords}

\section{Introduction}
\label{intro}

Large-scale structures in the Universe perturb the propagation of light rays from background sources causing small shape distortions and luminosity changes for their observed images \citep[e.g.,][]{Bartelmann2001}. Such effects, namely weak lensing effects, are closely related to the 
formation and evolution of foreground structures and the global expansion history of the Universe, and therefore are 
known to be one of the most promising probes in cosmological studies \citep[e.g.,][]{Albrecht2006, LSST2012, Euclid2012, Weinberg2013}. 
The cosmic shear two-point (2-pt) correlation analysis has been demonstrated to be a powerful statistics in extracting weak lensing signals 
from shape measurements of background galaxies \citep[e.g.,][]{Fu2008, Heymans2012, Kilbinger2013, Jee2013, Kitching2014}.
On the other hand, 2-pt correlations can only reveal part of the cosmological information embedded in weak lensing signals given the 
nonlinearity of the structure formation. To overcome this limitation, higher order cosmic shear correlation analyses are a natural extension
\citep[e.g.,][]{Semboloni2011, Waerbeke2013, Fu2014}. Weak lensing peak statistics, i.e., concentrating on high signal regions, is another way to probe efficiently 
the nonlinear regime of the structure formation, and thus can provide important complements to the cosmic shear 2-pt correlation analysis 
\citep[e.g.,][]{White2002, Hamana2004, Tang2005, Hennawi2005, Dietrich2010, Kratochvil2010, Yang2011, Marian2012, Hilbert2012,
Bard2013, Lin2015}

Observationally, different analyses have proved the feasibility of performing weak lensing peak searches from data
\citep[e.g.,][]{Wittman2006, Gavazzi2007, Miyazaki2007, Geller2010}. However, up to now, few cosmological constraints
are derived from weak lensing peak statistics in real observations. 
There are two main reasons for lack of such analyses. 
First, weak lensing observations are just starting to reach 
significantly large survey areas to provide reasonable statistics for peak abundances 
\citep[e.g.,][]{Shan2012, Waerbeke2013, Shan2014}. 
The second reason is the theoretical difficulty to calculate the cosmology dependence 
of peak abundances.

Theoretically, high Signal-to-Noise (S/N) weak lensing signal regions 
are expected to be associated closely with massive structures
along lines of sight \citep[e.g.][]{White2002}. 
Therefore, in principle, weak lensing peak abundances should reflect 
the underlying mass function of dark matter halos
weighted by the lensing efficiency kernel \citep[e.g.,][]{Hamana2004}. 
In practice, however, the correspondence between weak lensing peaks 
and the massive dark matter halos is influenced by various effects, 
such as the noise from the intrinsic ellipticities of 
source galaxies, the projection effect of large-scale structures, 
and the hierarchical mass distribution of dark matter halos 
\citep[e.g.,][]{Waerbeke2000, Tang2005, Hamana2012, Yang2013}. 
Thus, it is not straightforward to predict the cosmology dependence of 
weak lensing peak abundances. One possible solution is to create a large 
number of simulation templates for weak lensing peak statistics 
densely sampled in cosmological-parameter space. By comparing the 
observational measurements with the templates, we can derive 
cosmological constraints \citep[e.g.][hereafter LPH2015]{Dietrich2010, Liuj2015}. 
Considering the large number of cosmological parameters and
 different physical and observational effects, 
such an approach can be numerically expensive. 
Another efficient way is to build theoretical models, which take
into account the impact of the different effects.   

Based on simulation studies, \cite{Marian2009, Marian2010} developed
a phenomenological model for hierarchically detected weak lensing
peak abundances in which the 2-D peak mass function is scaled to 
the 3-D mass function of dark matter halos. \cite{Hamana2004} and \cite{Hamana2012} 
derived a fitting formula for weak lensing peak abundances by 
incorporating a probability function in relating peak heights 
and underlying dark matter halos at a given mass and redshift. 
Calibrated with numerical simulations, such a probability function 
tends to include the effects of noise from intrinsic ellipticities 
of source galaxies, the projection effects of large-scale structures 
and the non-spherical matter distributions of dark matter halos. 
Assuming Gaussian random fields for both the projected field of 
large-scale structures and the shape noise,  \cite{Maturi2010} 
proposed a theoretical model to calculate the number of contiguous areas above a given
threshold in the filtered convergence field. This is equivalent to the genus in Minkowski functionals. 
When the threshold is high, this statistics corresponds well to the number of peaks.
By comparing with simulations, it was shown that the model can predict well the 
number distribution for relatively low thresholds, but underestimates 
the high threshold regions that are mostly related to individual massive halos 
\citep{Maturi2010, Petri2013}.

In \citet[][hereafter F10]{Fan2010}, we have presented a theoretical 
model taking into account the shape noise effects. In this model, we divide a given area into 
halo regions occupied by dark matter halos with the size limited by their virial radii, 
and the regions outside dark matter halos. We first calculate
the weak lensing peak abundances in a halo region by assuming a density 
profile for the halo and the Gaussianity of the shape noise. 
By employing the mass function of dark matter halos with 
a lower mass cut representing the halo mass above which single 
halos dominantly contribute to weak lensing peaks along 
their lines of sight, we can then calculate statistically 
the peak distribution in regions occupied by massive halos. 
For the rest of the regions, we assume that the peaks are 
purely noise peaks. Our model has been tested extensively 
by comparing with numerical simulations \citep[F10;][hereafter LWPF2014]{Liu2014}.
It was shown that the model results are in very good agreement 
with simulation results. It is noted that the model in its present form
does not contain the projection effects of large-scale structures. 
For current generation of weak lensing surveys with 
the surface number density of lensing-usable galaxies 
around $n_g\sim 10\hbox{ arcmin}^{-2}$, the shape noise is dominant 
over the projection effects for the smoothing scale $\sim 1 \arcmin$. 
Thus neglecting the projection effects should not affect the model 
prediction significantly. For future surveys with much improved 
statistics, the projection effects need to be considered carefully, 
and we have started to look into this problem.    

For the CFHT Stripe 82 Survey 
\citep[CS82; e.g., ][]{Hand2013, Shan2014, Li2014}, the number density of galaxies 
used in weak lensing studies is $\sim 10\hbox{ arcmin}^{-2}$, 
and the survey area excluding the masked regions 
is $\sim 130\hbox{ deg}^2$. For this survey, the shape noise is the 
dominant source of contaminations on weak lensing peak analyses. 
We thus expect that our model can work well in predicting 
theoretically the peak abundances. This in turn allows us 
to perform cosmological constraints from observational weak lensing
peak abundances.  

As this work was being completed, we became aware of the study by LPH2015. 
They also analysed the cosmological application of weak lensing peak statistics using CFHTLenS data.
Their studies are based on interpolations from a suite of simulation templates on a grid of $91$ cosmological models
in the parameter space of $(\Omega_{\rm m}, \sigma_8, w)$ 
where $\Omega_{\rm m}$, $\sigma_8$ and $w$ are respectively the 
dimensionless matter density of the Universe, 
the amplitude of the extrapolated linear matter density fluctuations smoothed over a 
top-hat scale of $8h^{-1}\hbox{Mpc}$, and the equation of state of dark energy.
Using different and independent approaches, LPH2015 and our work both
showed the promising potential of weak lensing peak statistics in cosmological studies.

The paper is organised as follows. 
In Section 2, we describe briefly the CS82 survey.
In Section 3, we present the procedures of weak lensing peak analyses. 
In Section 4, we show the cosmological constraints derived from peak abundances.
Summary and discussion are given in Section 5.


\section{CFHT Stripe 82 Survey and Weak Lensing Catalogues} 
\label{data}

The CS82 survey was conducted thanks to the collaboration between the Canadian, French and Brazilian CFHT communities. CS82 covers a large fraction of the SDSS Stripe 82 with high quality 
$i$-band imaging under excellent seeing conditions in the range of $0.4$ to $0.8$ arcsec with an average of $0.59$ arcsec. 
The survey contains a total of $173$ tiles, $165$ of which from CS82 observations and 8 from CFHT-LS Wide \citep{Erben2013}. 
Each CS82 tile was obtained from 4 consecutive dithered observations 
each with an exposure time of 410 seconds. 
The derived $5\sigma$ limiting magnitude in a $2\arcsec$  diameter aperture is $i_{AB}\sim24$.
After removing overlapping regions and applying all the masks across the entire survey, the effective survey area is reduced from 
$\sim173\deg^2$ to $\sim130\deg^2$.

The same forward modelling {\it lens}fit pipeline \citep{Miller2007, Miller2013} as that for CFHTLenS was used for 
the CS82 shape measurements. As described in \cite{Miller2013}, the {\it lens}fit algorithm applied to CFHTLenS
was calibrated using different sets of simulated images with different observing conditions and PSFs. 
For the shape measurement errors written in the form of $\boldsymbol {\epsilon}=(1+m)\boldsymbol {\epsilon^{\rm true}}+\boldsymbol {c}$, 
it is found that the multiplicative bias factor $m$ can be well modelled as a function of galaxy signal-to-noise ratio and size. 
For the additive bias $\boldsymbol {c}$, the simulation calibration shows that it is consistent with zero. However for 
real data the additive bias can occur. \cite{Heymans2012} show that for the {\it lens}fit measurements of CFHTLenS, 
the $c_1$ component is consistent with zero, but there are small residues for $c_2$. Similarly to $m$, the $c_2$ term also
depends on galaxy signal-to-noise ratio and size. It is noted that CFHTLenS has a wide range of seeings, number of exposures, noise and depth,
and their influences on shape measurements can all be encoded into the two parameters of galaxy signal-to-noise ratio and size. 
Therefore for CS82, although the observing conditions are different from CFHTLenS, the {\it lens}fit pipeline is well applicable.
We should emphasise that although we use the same {\it lens}fit pipeline as CFHTLenS, 
we measure the PSFs and calculate the signal-to-noise ratio and size and the corresponding
bias terms $m$ and $c$ for CS82 source galaxies ourselves. 

In our weak lensing analyses, the selection criteria of source 
galaxies are weight $w>0$, FITCLASS$=0$, MASK$\le1$. Here the weight factor 
is the inverse variance weight accorded to each source galaxy given by {\it lens}fit. 
The FITCLASS is an index for star/galaxy classification provided by {\it lens}fit
with FITCLASS$=0$ for galaxies. The index MASK describes the mask information 
at an object's position. 
Objects with MASK $\le 1$ can safely be used for most weak lensing analyses \citep{Erben2013}. 
No magnitude cut is applied for the catalogue as fainter galaxies have lower weights. 
These criteria result in a total number of 
source galaxies of $9,281,681$. The total effective number of galaxies taking into account their weights is $5,475,318$, 
and the corresponding average effective number density is $\sim 11.8$ galaxies per arcmin$^2$. 

\begin{figure}
\centering
\includegraphics[width=0.95\columnwidth]{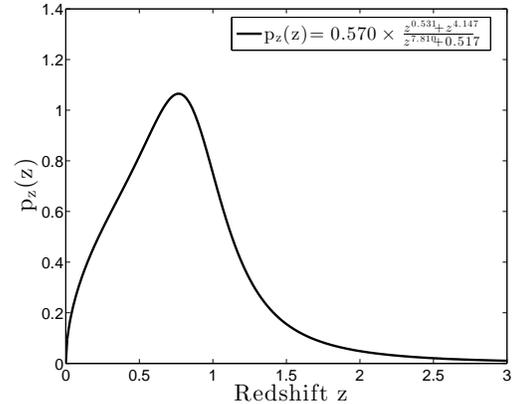}
\caption{Normalised redshift distribution of CS82 source galaxies.
}
\label{fig:pz_plot}
\end{figure}

Not all the source galaxies have redshift information. In our theoretical calculations, we therefore adopt a redshift distribution 
derived for the whole population of source galaxies, which is obtained by magnitude matching of COSMOS galaxies with the CS82 source galaxies 
\citep{Hand2013, Shan2014}. It is given by
\begin{equation}
p_z(z)\propto \frac{z^{a}+z^{ab}}{z^{b}+c},
\label{pz}
\end{equation}
where $a=0.531$, $b=7.810$ and $c=0.517$. The median redshift is $z_m=0.76$ and the mean redshift is $z=0.83$.
The normalised redshift distribution is shown in Figure~\ref{fig:pz_plot}. 

Because the COSMOS field is small, the sample variance can be significant. There are also errors in the photometric redshift estimations for COSMOS galaxies.
Thus the CS82 redshift distribution derived from COSMOS can have uncertainties. We will discuss the impact of such uncertainties on peak analyses in \S 4.3.

\section{Weak lensing peak analysis}
\label{peak analyses}

\subsection{Theoretical aspects}

In the weak lensing theory under the framework of general relativity, 
the deflection of light rays from a source can be written as
the gradient of a lensing potential $\psi(\boldsymbol \theta)$. 
The induced observational effects can be described by the 
Jacobian matrix $\boldsymbol A$, which is given by \citep[e.g.,][]{Bartelmann2001}
\begin{equation}
\boldsymbol{A}=\bigg (\delta_{ij}-\frac {\partial^2 \psi(\boldsymbol{\theta})}{\partial \theta_i\partial \theta_j} \bigg )=
\begin{pmatrix}
 1-\kappa-\gamma_1  & -\gamma_2 \\
 -\gamma_2 & 1-\kappa+\gamma_1
\end{pmatrix},
\label{JacobianMatrix}
\end{equation}
where the convergence $\kappa$ and the shear $\gamma_i$ lead to the isotropic change and the elliptical shape distortion of 
the observed image, respectively, with respect to the unlensed image. They are related to the potential by  
\begin{equation}
\kappa=\frac{1}{2}\nabla^2\psi, \quad
\gamma_1=\frac{1}{2}\bigg (\frac{\partial^2\psi}{\partial^2\theta_1}-\frac{\partial^2\psi}{\partial^2\theta_2}\bigg ), \quad
\gamma_2=\frac{\partial^2\psi}{\partial \theta_1 \partial \theta_2}.
\label{kappashear}
\end{equation}
In the weak lensing regime under the Born approximation, we have 
\begin{equation}
\kappa=\frac{3H_0^2\Omega_{\rm m}}{2c^2}\int_0^\chi d\chi'\frac{f_K(\chi')f_K(\chi-\chi')}{f_K(\chi)}\frac{\delta[f_K(\chi')\boldsymbol {\theta},\chi']}{a(\chi')},
\label{born}
\end{equation}
where $H_0$ is the Hubble constant, 
$\chi$ is the comoving radial distance, $f_K$ is the comoving angular diameter distance, $a$ is the scale factor of the universe,
and $\delta$ is the density perturbation along the line of sight.

The convergence $\kappa$ is directly related to the projection of line-of-sight density fluctuations weighted by the 
lensing efficiency factor. Physically, massive structures generate large weak lensing signals along their lines of sight. These peak signals are 
best seen visually in the weak lensing convergence field. On the other hand, weak lensing signals directly extracted from
galaxy shape measurements are the shears, or more precisely the reduced shears defined as $g_i=\gamma_i/(1-\kappa)$, 
rather than the convergence. Therefore weak lensing peak analysis usually involves procedures to construct quantities
representing the projected mass distribution from lensing shears
based on the relation between the two quantities shown in Eqn.(\ref{kappashear}). 

Specifically, assuming precise shape measurements, the observed ellipticity of a galaxy at redshift $z$ located at the sky position $\boldsymbol \theta$
can be written in a complex form given by \citep[e.g.,][]{Seitz1997} 
\begin{equation}
\boldsymbol {\epsilon}(\boldsymbol \theta, z)=\left\{ \begin{array}{ll} \frac{\boldsymbol{\epsilon}_s(\boldsymbol \theta, z)+
\boldsymbol{g}(\boldsymbol \theta, z)}{1+\boldsymbol{g^*}(\boldsymbol \theta, z)\boldsymbol{\epsilon}_s(\boldsymbol \theta, z)}
& \textrm{for $\vert{\boldsymbol{g}(\boldsymbol \theta,z)}\vert\leq 1$}\\ \\ \frac{1+\boldsymbol{g}(\boldsymbol \theta, z)
\boldsymbol{\epsilon}_s^{*}(\boldsymbol \theta, z)}{\boldsymbol{\epsilon}_s^{*}(\boldsymbol \theta, z)+\boldsymbol{g^*}(\boldsymbol \theta, z)}
& \textrm{for $\vert{\boldsymbol{g}(\boldsymbol \theta, z)}\vert>1$}
 \end{array}\right.
\label{eobs}
\end{equation}
where `*' represents the complex conjugate operation, $\boldsymbol g=g_1+ig_2$ is the complex reduced shear, 
and $\boldsymbol{\epsilon}_s$ is the intrinsic ellipticity of the galaxy.
Here the complex ellipticity is defined as $\boldsymbol \epsilon=(a-b)/(a+b)\exp(2i\phi)$ with $a$, $b$ and $\phi$ being the 
length of the major and minor axes and the orientation of the approximate ellipse of the observed image, respectively.
For source galaxies at a fixed redshift $z$, it has been shown that the average of $\boldsymbol \epsilon$ 
over a large number of galaxies near $\boldsymbol \theta$ gives rise to an unbiased estimate of $\boldsymbol {g}(\boldsymbol \theta, z)$ 
or $1/{\boldsymbol g}(\boldsymbol \theta, z)$ assuming $\langle \boldsymbol \epsilon_s\rangle=0$. For galaxies with a redshift distribution, 
the average of $\boldsymbol \epsilon$ over galaxies near a given sky position may have a complicated relation with the lensing signal
we are interested in if both $|\boldsymbol g|<1$ and $|\boldsymbol g|>1$ occur in the region for galaxies at different redshifts.
On the other hand, for sub-critical regions with $|\boldsymbol g|<1$ for all the redshifts, the average of the observed ellipticity
$\langle\boldsymbol \epsilon\rangle$ gives rise to an estimate of $\langle \boldsymbol g\rangle$  weighted by the redshift distribution of source galaxies.
In the case $\kappa\ll 1$, we have $\langle \boldsymbol \epsilon\rangle\approx \langle \boldsymbol \gamma\rangle $. For the weak lensing peak analysis, we
therefore need to construct a field closely related to the matter distribution from the (reduced) shear estimate $\langle \boldsymbol \epsilon\rangle$.

The aperture mass peak analysis, also referred to
as the shear peak analysis, is to study peaks in the aperture mass $\rm M_{\rm ap}$ field constructed from the tangential shear component with respect
to the point of interest with a filtering function $Q$ \citep[e.g.,][]{Schneider1998, Marian2012, Bard2013}. 
Theoretically, $\rm M_{\rm ap}$ corresponds to the convergence field filtered
with a compensated window function $U$ where $U$ and $Q$ are related. One of the advantages of $\rm M_{\rm ap}$ studies is that because of the
compensated nature of $U$, $\rm M_{\rm ap}$ is independent of the lensing mass-sheet degeneracy. Furthermore, in the case with $\kappa\ll 1$ and
$\boldsymbol g\approx \boldsymbol \gamma$, $\rm M_{\rm ap}$ can be obtained directly from the tangential component of the observed $\langle \epsilon_t\rangle $.
In the peak regions where the lensing signals are high, the difference between $\boldsymbol g$ and $\boldsymbol \gamma$
is not negligible. Therefore noting that the $\rm M_{\rm ap}$ constructed from $\langle \epsilon_t\rangle$ itself carries cosmological information,
it is not the same as the filtered convergence field. 

Another approach for weak lensing peak studies is to reconstruct the convergence field from $\langle \boldsymbol \epsilon\rangle$  
taking into account the nonlinear relation between $\boldsymbol g$ and $\boldsymbol \gamma$ \citep[e.g.,][]{Kaiser1993, Kaiser1995, 
Seitz1995, Bartelmann1995, Squires1996, Seitz1997, Jauzac2012, Jullo2014}. In this approach, to avoid unphysical
results, it is important to filter $\boldsymbol \epsilon$ first and then to proceed with convergence reconstruction using 
the filtered $\langle \boldsymbol \epsilon\rangle$. Different reconstruction schemes have been studied. For the classical Kaiser-Squires (KS) reconstruction
\citep{Kaiser1993, Squires1996}, boundary effects and the mass-sheet degeneracy problem can exist. 
However, for a field of view of about $1\deg\times 1\deg$ and larger, 
such effects are expected to be insignificant. In this paper, we reconstruct the convergence field
from the filtered $\langle \boldsymbol \epsilon\rangle $ with the nonlinear KS method  \citep[e.g.,][]{Bartelmann1995}. 
From numerical simulations, we find that the
regions with $|\boldsymbol g|>1$ are negligible, and therefore assuming sub-criticality for all the regions is an excellent 
approximation. 

\begin{figure*}
\includegraphics[width=0.45\textwidth]{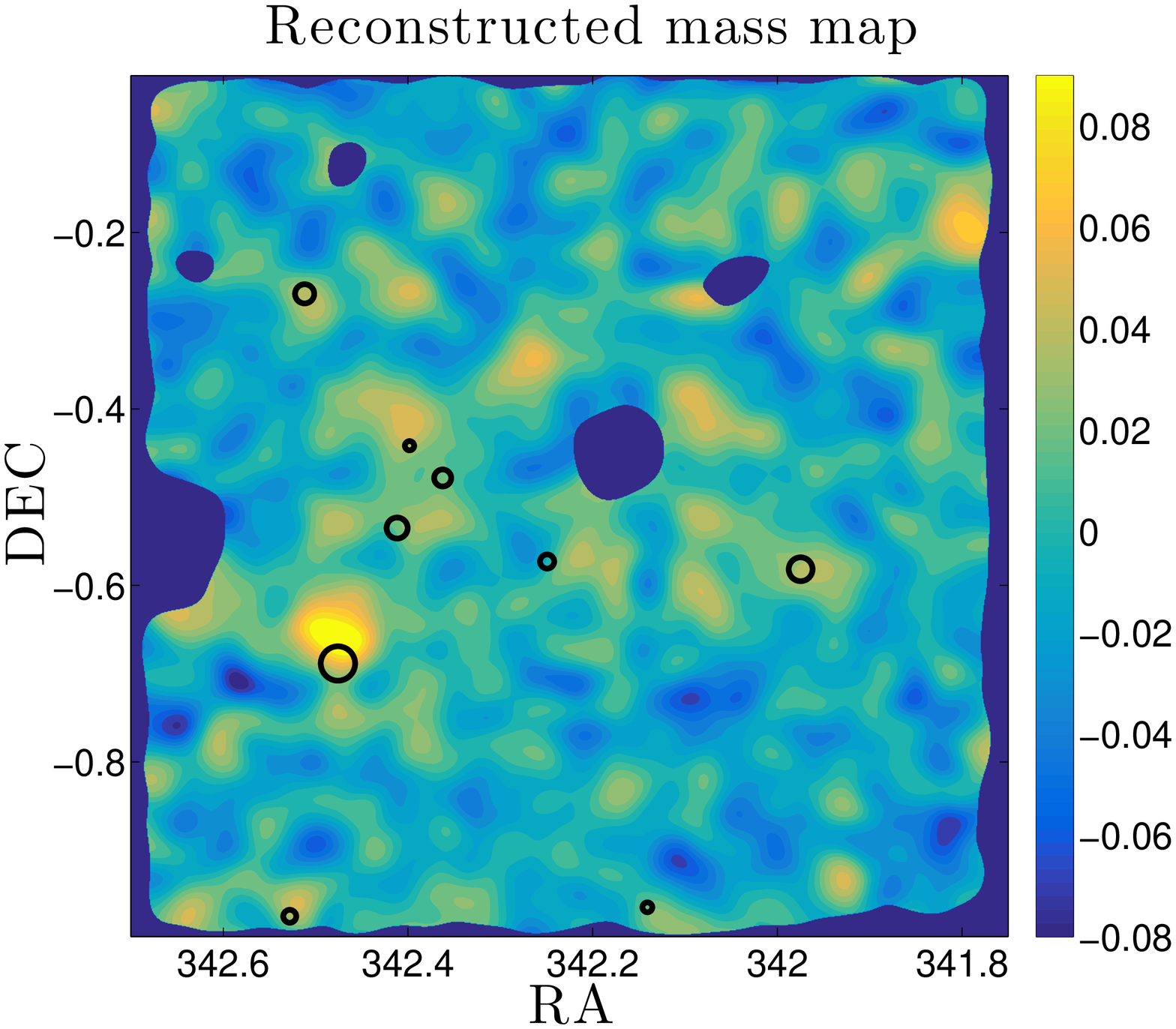}
\includegraphics[width=0.45\textwidth]{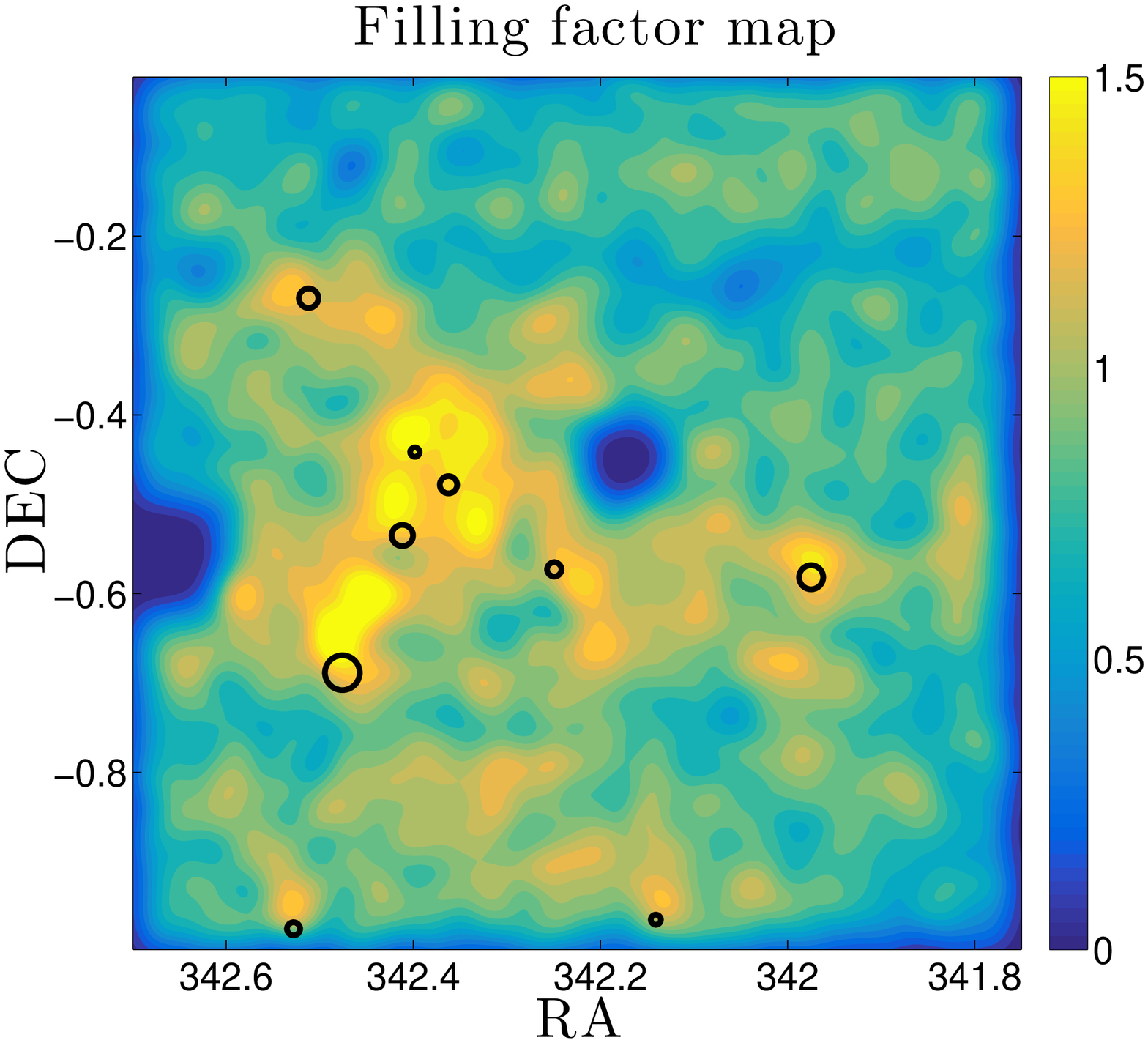}
\caption{Left panel: The reconstructed convergence map for one specific tile with the Gaussian smoothing scale $\theta_{\rm G}=1.5$ arcmin. 
Regions with filling factor $<0.5$ are masked out in dark blue. Right panel: The corresponding filling factor map. 
The redMaPPer clusters in the field are indicated by the black circles with the size indicating the richness of the clusters. 
} \label{fig:fillagrec}
\end{figure*}


\subsection{The convergence reconstruction and the peak identification}
\label{3_2}

Our convergence reconstruction procedures are described below.

For a source galaxy used in weak lensing analyses in the CS82 catalogue, we first correct the additive errors    

\begin{equation}
\epsilon_1^c=\epsilon_1, \quad
\epsilon_2^c=\epsilon_2-c_2,
\label{ecorrection}
\end{equation}
where $\epsilon_i$ and $\epsilon_i^c$ are the uncorrected and corrected ellipticity components, respectively, and $c_2$ is the 
additive bias given by CS82. With $\epsilon_i^c$, we apply smoothing and obtain a smoothed field of $\boldsymbol \epsilon$ 
on regular $1024\times1024$ grids over the field of view of one pointing. With the multiplicative errors $m$ taken into 
account statistically \citep{Waerbeke2013},  we have

\begin{equation}
\mathbf{\langle{\boldsymbol \epsilon}\rangle(\boldsymbol{\theta})}=\frac{\sum_{j}W_{\theta_{\rm G}}(\boldsymbol{\theta_j}-\boldsymbol{\theta})w(\boldsymbol{\theta_j})
\mathbf{\epsilon}^c(\boldsymbol{\theta_j})}{\sum_{j}W_{\theta_{\rm G}}(\boldsymbol{\theta_j}-\boldsymbol{\theta})w((\boldsymbol{\theta_j})(1+m_j)},
\label{smoothing}
\end{equation}
where $\boldsymbol \theta$ and $\boldsymbol \theta_j$ are for the grid position and the galaxy position, respectively,
$W_{\theta_{\rm G}}$ is the normalised smoothing function, $w$ is the weight for source galaxy shape measurements given by CS82. The summation is
over all the source galaxies. In the subcritical approximation, $\langle{\boldsymbol \epsilon}\rangle(\boldsymbol{\theta})$
is an unbiased estimate of $\langle \boldsymbol g\rangle$ smoothed over the window function $W_{\theta_{\rm G}}$ and weighted by the
source redshift distribution. We use the Gaussian smoothing function $W_{\theta_{\rm G}}$ with 

\begin{equation}
W_{\theta_{\rm G}}(\boldsymbol{\theta})=\frac{1}{\pi\theta_{\rm G}^2}\exp{\left(-\frac{|\boldsymbol{\theta}|^2}{\theta_{\rm G}^2}\right)}.
\label{window}
\end{equation}
The smoothing scale $\theta_{\rm G}$ is chosen to be $\theta_{\rm G}=1.5$ arcmin, suitable for cluster-scale structures that are
closely related to high weak lensing peaks. Within the smoothing kernel, the number of galaxies is about $n_g\theta_{\rm G}^2\sim 20$,
for $n_g\sim 10 \hbox{ arcmin}^{-2}$. We expect that the statistics of the residual shape noise 
after smoothing is approximately Gaussian from the central limit theorem \citep[e.g.][]{Waerbeke2000}.

With $\langle{\boldsymbol \epsilon}\rangle(\boldsymbol{\theta})$, we perform the convergence reconstruction iteratively by using the
relation between $\kappa $ and $\gamma$ in Eqn.~(\ref{kappashear}). Particularly, we use their relation in Fourier space with 

\begin{equation}
\hat \gamma(\boldsymbol k)=\pi^{-1}\hat D(\boldsymbol k)\hat \kappa (\boldsymbol k),
\label{kgfourier}
\end{equation}
and 
\begin{equation}
\hat D(\boldsymbol k)=\pi\frac{k_1^2-k_2^2+2ik_1k_2}{k_1^2+k_2^2}.
\label{dfourier}
\end{equation} 

We start by assuming $\kappa^{(0)}=0$ everywhere, and thus $\boldsymbol \gamma^{(0)}=\langle\boldsymbol \epsilon\rangle$ 
\citep{Bartelmann1995}.
At ${n}$-th step, we obtain $\kappa^{(n)}$ from $\boldsymbol{\gamma}^{(n-1)}$ via Eqn.(\ref{kgfourier}) and the subsequent
inverse Fourier transformation. We then update $\boldsymbol \gamma$ to $\boldsymbol{\gamma}^{(n)}=(1-\kappa^{(n)})\langle\boldsymbol \epsilon\rangle$
for next iteration. The reconstruction process is stopped when the converging accuracy of $10^{-6}$ 
(the maximum difference of the reconstructed convergence between the two sequential iterations) is reached. 
For CS82, the reconstruction is done pointing by pointing each with the field of view of about $1\deg\times 1\deg$.

To evaluate the shape noise level in each pointing for subsequent peak analyses, we randomly rotate the corrected 
ellipticity of each galaxy. Then the same procedure is applied to obtain the reconstructed random noise convergence field.

It is noted that there are regions with no reliable shape measurements for galaxies indicated with the index MASK$>$1 in the CS82 catalogue. 
These galaxies are excluded in the weak lensing analyses. The existence of these masked regions can affect the weak lensing peak abundances
significantly if they are not treated properly. In LWPF2014, we study in detail the mask effects. With the signal-to-noise ratio defined
by the average noise level, the number of high peaks increases around the masked regions, which can lead to 
considerable bias in cosmological parameter constraints derived from weak lensing peak abundances. 
To reduce the mask effects, regions around masks should be excluded in peak counting. If they are kept, the noise effects
in these regions should be considered separately from the regions away from masks (LWPF2014).

To quantify the mask effects on the number of usable galaxies in the convergence reconstruction, we calculate the galaxy filling factor
at each grid point similar to that done in \cite{Waerbeke2013}. Summing over galaxies outside masked regions, we define the
galaxy filling factor as

\begin{equation}
f(\boldsymbol{\theta})=\frac{\sum_{j}W_{\theta_{\rm G}}(\boldsymbol{\theta_j}-\boldsymbol{\theta})w(\boldsymbol{\theta_j})}{f_0},
\label{filling}
\end{equation}
where $f_0$ is calculated by randomly populating galaxies over the full area of a tile with 

\begin{equation}
f_0=\langle \sum_{n}W_{\theta_{\rm G}}(\boldsymbol{\theta_n}-\boldsymbol{\theta})
\tilde {w}(\boldsymbol{\theta_n})\rangle.
\label{f0}
\end{equation}
Here $\langle \rangle$ is for the average over $\boldsymbol \theta$.
Specifically, for each tile, we calculate the average number density of galaxies in the area excluding the masked regions. 
With this number density, we then randomly populate galaxies over the full area of the tile including the masked regions.
For each galaxy, we randomly assign it a weight $\tilde {w}$ according to the weight distribution of the real observed galaxies.
We then calculate the quantity $\sum_{n}W_{\theta_{\rm G}}(\boldsymbol{\theta_n}-\boldsymbol{\theta})
\tilde {w}(\boldsymbol{\theta_n})$ at each grid point $\boldsymbol \theta$ where the summation is over all the 
populated galaxies. The average value over all the grid points gives rise to $f_0$. 

Hence, for each pointing, we obtain the reconstructed lensing convergence, noise and filling factor maps, respectively.
We have a total of $173$ sets of maps corresponding to the $173$ pointings.
In Figure~\ref{fig:fillagrec}, we show an example of the reconstructed convergence map for one pointing and 
the corresponding map of the filling factor. In each map, the dark blue regions are regions with the filling factor $f<0.5$.
To avoid the mask effects on weak lensing peak analyses, we exclude these regions in peak counting \citep[LWPF2014;][]{Waerbeke2013}. 
The black circles in the plots show the clusters in the field detected using the red-sequence Matched-filter Probabilistic Percolation (redMaPPer)
algorithm \citep{Rykoff2014}. It is seen that the clusters have a good
association with weak lensing convergence peaks, but the correspondence is not one to one due to the existence of noise and the projection effects
of large-scale structures \citep[e.g.,][]{Shan2014}.

For weak lensing peak analyses, we detect peaks from the reconstructed lensing convergence maps as follows.
Considering a pixel on a reconstructed convergence map ($1024\times1024$ pixels), if its convergence value is the highest among 
its nearest 8 neighbouring pixels, it is identified as a peak. We only count peaks in regions with the filling factor $f> 0.5$.
To reduce the boundary effects, we also exclude the outer most 50 pixels (corresponding to $\sim 3\hbox{ arcmin}\sim 2\theta_{\rm G}$) 
in each of the four sides of a map in our peak counting. The effective area of CS82 for weak lensing peak studies is then reduced to 
$\sim 114\deg^2$ after mask-region and boundary exclusions. The signal-to-noise ratio of a peak is defined by

\begin{equation}
\nu=\frac{K}{\sigma_0},
\label{snr}
\end{equation}
where $K$ is the reconstructed convergence value of the peak and $\sigma_0$ is the mean rms of the noise from the $173$ noise maps. 
It is known that $\sigma_0$ depends on the number density of source galaxies and the smoothing scale of the window function used in 
obtaining the smoothed ellipticity field $\langle\epsilon\rangle$. It can vary somewhat from one pointing to another. 
In our study, the mean $\sigma_0$ is evaluated from all the noise maps considering only regions with the filling factor $f>0.5$. 
For the smoothing scale $\theta_{\rm G}=1.5$ arcmin, we have $\sigma_0\approx0.022$.
It is noted that we calculate $\sigma_0$ directly from the rotated galaxies in noise maps,
and therefore we do not need to know explicitly the galaxy intrinsic ellipticity dispersion $\sigma_{\epsilon}$. On the other hand,
we find that for CS82 galaxies, $\sigma_{\epsilon}\sim 0.4$ for the total of the two components.

We note that in our peak identification scheme, we do not group peaks together as some other methods do \citep[e.g.,][]{Hamana2012}. 
We will see in \S 3.3 that our theoretical model for peak abundances takes into account the noise peaks in halo regions and counts them as independent ones.  
Accordingly, we therefore do not need the peak grouping that may give rise to some artificial effects. 


\subsection{Theoretical model for weak lensing peak abundances}

To derive cosmological constraints from observed weak lensing peak abundances, their dependence on cosmological models needs to be understood
and quantified. Known to be closely related to line-of-sight matter concentrations, the existence of galaxy shape noise, the projection effects
of large-scale structures, the complex mass distribution of dark matter halos, etc., complicates the relation between weak lensing peak abundances 
and the underlying mass function of dark matter halos. While building a large set of templates from numerical simulations densely sampled over
the multi-dimensional cosmological parameter space can be very useful and important, it is computationally expensive. On the other hand, 
theoretical modelling based on our physical understandings can be very insightful and valuable in disentangling 
different effects on weak lensing peak abundances. With explicit dependences on cosmological parameters and other physical parameters,
it can be used to perform cosmological constraints efficiently. Depending on the assumptions and approximations employed in the modelling,
their results can be less accurate than those from full simulations. Similarly to the calculation of the nonlinear matter power spectrum 
based on the halo model but calibrated with simulations, which is widely used in weak lensing two-point correlation analyses, 
the combination of the two, that is, testing and calibrating a model with simulations, can be a very effective way to study 
the cosmological dependence of weak lensing peak statistics efficiently with high precision. 
 
In this paper, we focus mainly on high peaks and adopt the model of F10 for the weak lensing peak abundances, which takes into account the dominant shape noise in the 
modelling. The model has been tested extensively by comparing with full ray-tracing simulations (F10, LWPF2014), and has also be
confronted with observational studies \citep{Shan2012,Shan2014}. 

Here we describe the important ingredients of the model. More details can be found in F10 and LWPF2014.

In F10, we assume that the smoothed convergence field can be written as $K_N = K + N$, where $K$ represents the true lensing convergence, 
and $N$ is for the residual shape noise. The field $N$ results from the contribution of the intrinsic ellipticities of different galaxies in 
the smoothing kernel. Without considering the intrinsic alignments of source galaxies, if the number of galaxies within the smoothing kernel
is large enough, it has been shown that $N$ is approximately a Gaussian random field from the central limit theorem (e.g., \citealt{Waerbeke2000}). 
As discussed in \S 3.2, for a smoothing scale $\theta_{\rm G}=1.5 \hbox{ arcmin}$, the number of galaxies within the smoothing window in CS82 is $\sim 20$. Therefore $N$ can be well approximated
as a Gaussian random field. 

Concentrating on high peaks, it is expected that signals of true peaks mainly come from individual massive halos 
\citep[e.g.,][]{Hamana2004, Tang2005, Yang2011}. We therefore divide a given area into halo regions and field regions. 
Inside the region of an individual halo, we have $K_N=K+N$, where $K$ is regarded as a known field from the halo convergence,
and $N$ is a Gaussian random field. Therefore $K_N$ itself is also a Gaussian random field modulated by the halo surface mass distribution $K$.
Then the peak number distribution for $K_N$ is readily calculable using Gaussian statistics. The modulation effects from $K$ involve $K$ itself, 
its first derivatives $K^i=\partial K/\partial x_i, (i=1,2)$ and its second derivatives $K^{ij}=\partial^2 K/\partial x_i\partial x_j$.  
The total number of peaks in halo regions can thus be obtained by the summation of the peaks over all the halo regions weighted by the halo mass function. 
In the field region, the numbers of peaks are directly computed from the noise field $N$. 

Specifically, the total surface number density of peaks can then be written as

\begin{equation}
n_{\mathrm{peak}}(\nu)d\nu=n_{\mathrm{peak}}^c(\nu)d\nu+n_{\mathrm{peak}}^n(\nu)d\nu,
\label{npeaktwoterm}
\end{equation}
where $\nu=K_N/\sigma_0$ is the signal-to-noise ratio of a peak. The term $n_{\mathrm{peak}}^c(\nu)$ is for peaks in halo regions 
including not only the true peaks corresponding to real halos but also the noise peaks within the halo regions. The second term
$n_{\mathrm{peak}}^n(\nu)$ is for pure noise peaks in field regions.

For $n_{\rm peak}^c(\nu)$ in halo regions, it can be written as
\begin{equation}
 n_{\mathrm{peak}}^c(\nu)=\int{dz\frac{dV(z)}{dzd\Omega}}\int_{\rm M_{\rm lim}}{dMn(M,z)f_p(\nu,M,z)},
\label{npeakc}
\end{equation}
where $dV(z)$ is the cosmological volume element at redshift z, $d\Omega$ is the solid angle element, $n(M,z)$ is the mass function of dark matter halos.
Here we adopt the Sheth-Tormen mass function in the calculation \citep{Sheth1999}. The mass limit $\rm M_{\rm lim}$ is for the mass above which
individual halos contribute dominantly to the weak lensing peak signals along their lines of sight. From simulation analyses, 
we find that $\rm M_{\rm lim}=10^{13.7}h^{-1}\hbox{M}_{\odot}$ is a suitable choice. The factor $f_p$ is for the 
number of peaks in the area within the virial radius of a halo of mass $M$ at redshift $z$, and is given by  
\begin{equation}
f_p(\nu,M,z)=\int_{0}^{\theta_{\mathrm{vir}}}d\theta\hbox{ } (2\pi \theta)\hbox{ } \hat {n}^c_{\mathrm{peak}}(\nu,\theta,M,z)
\label{fnumz}
\end{equation}
where $\theta_{\rm vir}=R_{\rm vir}(M,z)/D_A(z)$, and $D_A(z)$ is the angular-diameter distance. The physical virial radius is calculated by

\begin{equation}
R_{\rm vir}(M,z)=\bigg[\frac{3M}{4\pi\rho(z)\Delta_{\rm vir}(z)}\bigg]^{1/3},
\label{virialr}
\end{equation}
where $\rho(z)$ is the background matter density of the Universe at redshift $z$ and
the overdensity $\Delta_{\rm vir}$ is taken from \cite{Henry2000}.

The function $\hat {n}^c_{\mathrm{peak}}(\nu,\theta,M,z)$ in Eqn.(\ref{fnumz}) describes 
the surface number density of peaks at the location of $\theta$ from the centre of
the halo, which depends on the convergence profile of the halo. On the basis of the theory of Gaussian random fields, 
it can be derived explicitly as (F10)
\begin{eqnarray}
&&\hat n^c_{\mathrm{peak}}(\nu,\theta, M, z)=\exp \bigg [-\frac{(K^1)^2+(K^2)^2}{\sigma_1^2}\bigg ]\nonumber \\
&&\times \bigg [ \frac{1}{2\pi\theta_*^2}\frac{1}{(2\pi)^{1/2}}\bigg ]
\exp\bigg [-\frac{1}{2}\bigg ( \nu-\frac{K}{\sigma_0}\bigg )^2\bigg ] \nonumber \\
&&\times \int_0^{\infty} dx_N\bigg \{ \frac{1}{ [2\pi(1-\gamma_N^2)]^{1/2}}\nonumber \\
&&\times \exp\bigg [-\frac{ [{x_N+(K^{11}+K^{22})/ \sigma_2
-\gamma_N(\nu_0-K/\sigma_0)}]^2}{ 2(1-\gamma_N^2)}\bigg ] \nonumber \\
&& \times  F(x_N)\bigg \},
\label{nchat}
\end{eqnarray}
where $\theta_*^2=2\sigma_1^2/\sigma_2^2$, $\gamma_N=\sigma_1^2/(\sigma_0\sigma_2)$, $K^{i}=\partial_i K$, 
and $K^{ij}=\partial_{ij} K$. Here the quantities $\sigma_i$ are the moments of the noise field $N$ given by (e.g. \citealt{Waerbeke2000})
\begin{equation}
\sigma_i^2=\int {d\boldsymbol k}\hbox{ }k^{2i}\langle |N(k)|^2\rangle,
\label{noisesigma}
\end{equation}
where $N(k)$ is the Fourier transform of the noise field $N$. For $K(\theta)$, $K^i(\theta)$, and $K^{ij}(\theta)$ of a halo with mass $M$ at redshift $z$, 
they depend on the mass profile of the halo and source redshift distribution. Here we assume the spherical Navarro-Frenk-White (NFW) 
mass distribution for dark matter halos \citep{NFW1996, NFW1997}. The concentration parameter $c_{\rm vir}(M)=R_{\rm vir}/r_{\rm s}$ is calculated from the 
mass-concentration relation given in \cite{Bhattacharya2013} where $r_s$ is the characteristic scale of an NFW halo. 
For the redshift distribution of source galaxies, we take Eqn. (\ref{pz}) for CS82 data.

The function $F(x_N)$ in Eqn.~(\ref{nchat}) is given by (F10)
\begin{eqnarray}
&&F(x_N)= \exp\bigg [-\frac{(K^{11}-K^{22})^2}{\sigma_2^2}\bigg ]
\times \nonumber \\
&& \int_0^{1/2}de_N \hbox{ }8(x_N^2e_N)x_N^2(1-4e_N^2) \exp(-4x_N^2e_N^2)
\times \nonumber \\
&& \int_0^{\pi} \frac{d\theta_N}{ \pi} \hbox{ }
\exp\bigg [-4x_Ne_N\cos (2\theta_N)\frac{(K^{11}-K^{22})}{\sigma_2}\bigg ]. \nonumber \\
\label{fxn}
\end{eqnarray}
where $x_N=(\lambda_{N1}+\lambda_{N2})/\sigma_2$ and $e_N=(\lambda_{N1}-\lambda_{N2})/(2\sigma_2 x_N)$. $\lambda_{N1}$ and $\lambda_{N2}$ 
are the two eigenvalues $(\lambda_{N1}\ge\lambda_{N2})$ and $\theta_N$ is the rotation angle in the range $[0,\pi]$ with the 
diagonalisation of $(-K_N^{ij})$ \citep[e.g.,][]{BBKS1986, BE1987}.

As for the field term $n_{\mathrm{peak}}^n(\nu)$ in Eqn.~(\ref{npeaktwoterm}), it is given by
\begin{equation}
 \begin{split}
n_{\mathrm{peak}}^{n}(\nu)=\frac{1}{d\Omega}\Big\{n_{\mathrm{ran}}(\nu)\Big[d\Omega-\int dz\frac{dV(z)}{dz}\\
   \times\int_{\rm M_{\rm lim}} dM\,n(M,z)\,(\pi \theta_{\rm vir}^{2})\Big]\Big\},
\end{split}
\label{npeakn}
\end{equation}
where $n_{\mathrm{ran}}(\nu)$ is the surface number density of pure noise peaks without foreground halos. 
It can be calculated by Eqn.~(\ref{nchat}) with $K=0$, $K^{i}=0$ and $K^{ij}=0$.

It is seen that in this model, the cosmological information is contained in the halo mass function, lensing kernel, 
cosmic volume element, and the density profile of dark matter halos. We note that although we use the NFW density profile 
and the mean mass-concentration relation derived by \cite{Bhattacharya2013} for the full sample of dark matter halos in our fiducial model calculations,  
in principle, the density profile parameters can be treated as free parameters. Therefore from weak lensing peak abundances, 
it is possible to constrain these structural parameters simultaneously with cosmological parameters. 

For our model calculation of $n_{\rm peak}$, multi-dimensional integrations are needed. 
With great efforts numerically and applying multiple parallel techniques, such as OpenMP and GPU programming, 
we have developed a fast and high precision model calculation algorithm, which makes it possible for us to perform
cosmological constraints from weak lensing peak abundances. An outline of our programming structures is given in the Appendix.


\section{Cosmological Constraints from CS82 Weak Lensing Peak Counts}
\label{constraints}

\subsection{Fitting method}

As described in \S 3.2, from the reconstructed convergence maps, we identify and count peaks in regions where the 
galaxy filling factor is $f>0.5$. This effectively excludes the masked regions in the peak counting to avoid the mask effects. 
The useful survey area is approximately $114\deg^2$. We only consider high peaks with the signal-to-noise ratio $\nu\ge 3$. 
With the noise level $\sigma_0\sim 0.022$ under the smoothing scale $\theta_G=1.5\hbox{ arcmin}$ for CS82, the high peaks 
have smoothed signals $K\ge 0.066$. For the considered survey area, there are few peaks with $\nu>6$. We therefore
concentrate on the peaks in the range of $3\le \nu \le 6$.

We divide the peaks into $5$ bins. We consider both equal bins with $\Delta \nu=0.5$ and unequal bins 
with the number of peaks comparable in different bins. For equal bins, we do not include the number of peaks in the bin of $\nu=(5.5, 6]$ 
in cosmological studies because it is only $\sim 1$ with large expected statistical fluctuations. In this case, the number of peaks is
$\sim 500$ for the first bin with $\nu=[3,3.5]$ and is $\sim 10$ for the last bin with $\nu=(5,5.5]$. For  
unequal bins, we have the number of peaks in the range of $(\sim 160, \sim 80)$ for different bins with
$\nu=[3, 3.1], (3.1, 3.25], (3.25, 3.5], (3.5, 4], (4, 6]$, respectively. 

To derive cosmological parameter constraints from weak lensing peak counts, we calculate the $\chi^2$ defined as follows (LWPF2014) 

\begin{equation}
 \chi_{p'}^{2}=\boldsymbol{dN}^{(p')}(\widehat{\boldsymbol{C}^{-1}})\boldsymbol{dN}^{(p')}
=\sum_{ij=1,...,5}dN_{i}^{(p')}(\widehat{C_{ij}^{-1}})dN_{j}^{(p')},
\label{chi2}
\end{equation}
where $dN_i^{(p')}=N_{\rm peak}^{(p')}(\nu_i)- N_{\rm peak}^{(d)}(\nu_i)$ with $N_{\rm peak}^{(p')}(\nu_i)$ being the prediction 
for the cosmological model $p'$ from F10 and $N_{\rm peak}^{(d)}(\nu_i)$ being the observed data for the peak counts. This 
effectively assumes that the number fluctuation in each bin can be approximated by a Gaussian distribution. With the number of
peaks in the equal bin case being larger than $\sim 10$ per bin and being larger than about $80$ per bin in the unequal 
bin case, the Gaussian error distribution is expected to be a good approximation. The corresponding likelihood function is given by 

\begin{equation}
 L\propto \exp\left(-\frac{1}{2}\chi_{p'}^2\right).
\label{likelihood}
\end{equation}

The matrix $C_{ij}$ is the covariance matrix of the peak counts including the error correlations between different $\nu$ bins.
Here we apply bootstrap analyses using the CS82 observational data themselves to obtain an estimate of $C_{ij}$.
We realise that such an estimate cannot reveal the cosmic variance over the full survey area.
Ideally, $C_{ij}$ should be constructed by generating a large number of CS82 mocks from ray-tracing simulations following exactly
the same galaxy distribution, mask distribution, survey geometry, etc. as CS82 data. It is noted that the CS82 survey 
covers a long stripe of $\sim 90\deg\times 2\deg$. Therefore to fully mimic the survey geometry, we need very large simulations to
cover the $\sim 90\deg$ extension, which are difficult to achieve at the moment. 

For $C_{ij}$, we thus generate $10 000$ bootstrap samples by resampling the 173 tiles from real observation data sets. 
The covariance matrix $C_{ij}$ is then calculated from the bootstrap samples by
\begin{equation}
 C_{ij}=\frac{1}{R-1}\sum_{r=1}^{R}[N^r_{\rm peak}(\nu_i)-{N}_{\rm peak}^{(d)}(\nu_i)][N^r_{\rm peak}(\nu_j)-{N}_{\rm peak}^{(d)}(\nu_j)],
\label{covar}
\end{equation}
where $r$ denotes for different samples with the total number of samples $R=10 000$, and $N^r_{\rm peak}(\nu_i)$ is for the
peak count in the bin centred on $\nu_i$ from the sample $r$. 
The inverse of the covariance matrix is then calculated by \cite{Hartlap2007}
\begin{equation}
\widehat{\boldsymbol{C}^{-1}}=\frac{R-N_{\mathrm{bin}}-2}{R-1}(\boldsymbol{C}^{-1}),~~N_{\mathrm{bin}}<R-2
\label{nobiascov}
\end{equation}
where $N_{\mathrm{bin}}$ is the number of bins used for peak counting.

We note that such bootstrap analyses implicitly assume the independence of the peak distribution 
between different tiles each with an area of about $1\deg^2$. We have made a test by dividing the survey area into units each
containing $4$ adjacent tiles (i.e., $\sim 2\deg\times 2\deg$) and then performing bootstrap analyses by resampling these units.  
The resulted inverse covariance, considering the diagonal elements which are much larger than the off-diagonal terms, shows
less than $\sim 8\%$ differences from the one using one tile as a unit. We also perform tests using $2 \times 3$ and $2 \times 10$
adjacent tiles as units, separately. The diagonal terms of the inverse covariance differ from those using one tile as a unit 
by $\le 8\%$ and $\le 11\%$, respectively. We have carried out another test to use our 15 sets of mocked data to calculate the covariance. 
However, our mock simulations do not mimic the long stripe geometry of the CS82 survey, 
and the independent unit has an area of $4\times 3.5\times 3.5=49\deg^2$, as will be described in \S 4.2. We find that for the diagonal elements of the inverse covariance, 
the differences between the results from mock sets and that of using one tile as a unit are generally less than $10\%$.

We also carry out Jackknife resampling for error estimations. For using one tile, $2\times 2$ and $2 \times 3$ tiles as independent
units, respectively, the diagonal elements of the resulted inverse covariance differ from the corresponding bootstrap resampling by less than $5\%$.

We therefore adopt the bootstrap covariance estimated from CS82 data using one tile as an independent unit in our following fiducial analyses. 

We use CosmoMC \citep{Lewis2002} modified to include our likelihood function for weak lensing peak counts to perform cosmological constraints.
In this paper, we mainly consider constraints on the two cosmological parameters $(\Omega_{\rm m}, \sigma_8)$ 
under the flat $\Lambda$CDM assumption. We adopt flat priors in the range of $[0.05, 0.95]$ and $[0.2, 1.6]$ for
$\Omega_{\rm m}$ and $\sigma_8$, respectively. 
We take the Hubble constant $h=0.7$ in units of $100\hbox{ }\rm {km}/\rm {s}/\rm{Mpc}$, 
the power index of the initial density perturbation spectrum $n_s=0.96$, and the present baryonic matter density $\Omega_{\rm b}=0.046$. 
As discussed above, our fiducial constraints use the covariance estimated from bootstrap resampling of individual tiles.
As a quantitative comparison, we also perform $(\Omega_{\rm m}, \sigma_8)$ constraints by using the covariance
estimated from resampling $2 \times 2$ adjacent tiles. The constraint contours are nearly overlapped with those of our fiducial analyses
with the area of $1\sigma$ region larger only by $\sim 1\%$.

To further show the potential of weak lensing peak statistics, we also perform constraints on the mass-concentration relation of dark matter halos, 
assuming it follows a power law relation, simultaneously with the cosmological parameters $(\Omega_{\rm m}, \sigma_8)$.
  
\subsection{Mock CS82 analyses}
\label{4_2}

Before presenting the results from CS82 observational data, in this part, we first show our mock CS82 analyses using ray-tracing simulations. 
As discussed above, the full mock of CS82 taking into account the long stripe geometry demands very large simulations, which are
yet to be realised. Therefore in the current mock analyses presented here, we do not attempt to mimic the
survey geometry of CS82, but are limited to the tests of our peak analyses procedures,
including the convergence reconstruction, mask effects exclusion, cosmological parameter fitting, etc.. 
These mocks also serve as a further test of our theoretical model on the peak abundances. 

To construct the CS82 mocks, we carry out dark-matter-only N-body simulations in the flat $\Lambda$CDM framework. The cosmological 
parameters are taken to be $\Omega_{\rm m}=0.28$, $\Omega_\Lambda=0.72$, $\Omega_{\rm b}=0.046$, $\sigma_8=0.82$, $n_{\rm s}=0.96$ and $h=0.7$.
Our ray-tracing procedures largely follow our previous studies of LWPF2014, but with different box paddings.

In accord with the redshift distribution of CS82 galaxies, we perform ray-tracing calculations up to $z=3$.  
For our considered cosmological model, the comoving distance to $z_s=3$ is approximately $4.54 h^{-1} \mathrm{Gpc}$. 
We design to pad $12$ independent simulation boxes to $z=3$, with $8$ simulation boxes each with a size of $320 h^{-1} \mathrm{Mpc}$ to redshift $z=1$
and $4$ larger simulations each with a size of $600 h^{-1} \mathrm{Mpc}$ from $z=1$ to $z=3$. 
With these independent simulations, we can perform ray-tracing calculations straightforwardly with no repeated use of 
same structures.  

The N-body simulations are done with Gadget-2 (Springel 2005). For both small and large sized simulations, we use $640^3$ particles. 
The mass resolution is $\sim 9.7\times10^9 h^{-1} \rm {M}_\odot$ and $\sim 6.4\times10^{10} h^{-1} \rm{M}_\odot$ for small boxes and large boxes, respectively. 
The simulations start from $z=50$ and the initial conditions are set by using 2LPTic (Crocce et al. 2006). The initial density perturbation spectrum
is generated by CAMB (Lewis et al. 2000). The force softening length is about $\sim20h^{-1} \mathrm{kpc}$, 
which is good enough for our studies concerning mainly high weak lensing peaks corresponding to massive dark matter halos. 

For ray-tracing simulations, we use $59$ lens planes up to $z=3$. The corresponding redshifts of the planes are listed in Table \ref{tab:zl}.
We follow closely the method of \cite{Hilbert2009}. The detailed descriptions for ray-tracing calculations can be found in LWPF2014.
In order to generate mock data for CS82 galaxies, we calculate shear and convergence maps at the far edge of each of the $59$ lensing planes
using the lower redshift planes. For a set of simulations with $12$ independent boxes, we then can generate $4$ sets of lensing maps each with 
an area of $3.5\times3.5\deg^2$ sampled on $1024\times 1024$ pixels. In each set, we have $59$ shear and $59$ convergence maps at $59$ 
different redshifts corresponding to the far edges of the $59$ lens planes.   
We run $12$ sets of simulations, and generate lensing maps with the total area of $12\times 4\times (3.5\times 3.5)=588\deg^2$. 
This allows us to generate $3$ nearly independent mocks for CS82, and each mock
is constructed from $3$ sets of simulations with the area of $\sim 3\times 49=147 \deg^2$. 

\begin{table}
\caption{Redshifts of the lens planes. The planes at $z_l>1.0$ are produced from $4$ independent $L=600 h^{-1}\mathrm{Mpc}$ simulations, 
while those at lower $z_l$ are obtained from $8$ independent simulations with $L=320 h^{-1}\mathrm{Mpc}$.}
\label{tab:zl}
\begin{center}
  \leavevmode
    \begin{tabular}{c c c c c c} \hline
        0.0107  & 0.0322 & 0.0540 & 0.0759 & 0.0981 & 0.1205 \\
        0.1432  & 0.1661 & 0.1893 & 0.2127 & 0.2364 & 0.2604\\
        0.2847  & 0.3094 & 0.3343 & 0.3596 & 0.3853 & 0.4113\\
        0.4377  & 0.4645 & 0.4917 & 0.5193 & 0.5474 & 0.5759\\
        0.6049  & 0.6344 & 0.6645 & 0.6950 & 0.7261 & 0.7578\\
        0.7900  & 0.8229 & 0.8564 & 0.8906 & 0.9254 & 0.9610\\
        0.9895  & 1.0289 & 1.0882 & 1.1496 & 1.2131 & 1.2789\\
        1.3472  & 1.4180 & 1.4915 & 1.5680 & 1.6475 & 1.7303\\
        1.8166  & 1.9066 & 2.0005 & 2.0987 & 2.2013 & 2.3087\\
        2.4213  & 2.5393 & 2.6632 & 2.7934 & 2.9296 & \\
        \hline
     
     \end{tabular}
    \end{center}
\end{table}

\begin{figure*}
\includegraphics[width=0.46\textwidth]{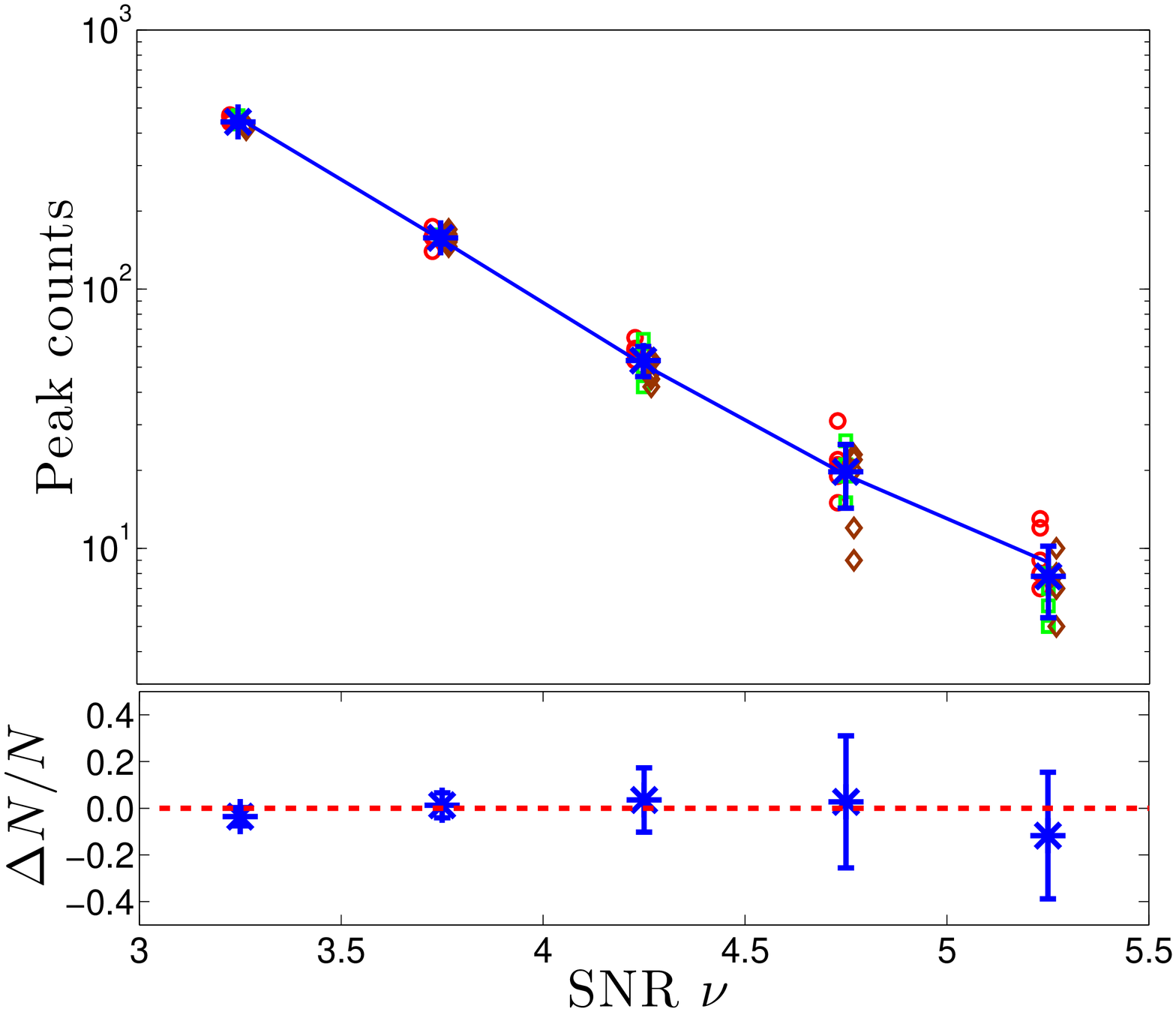}
\includegraphics[width=0.46\textwidth]{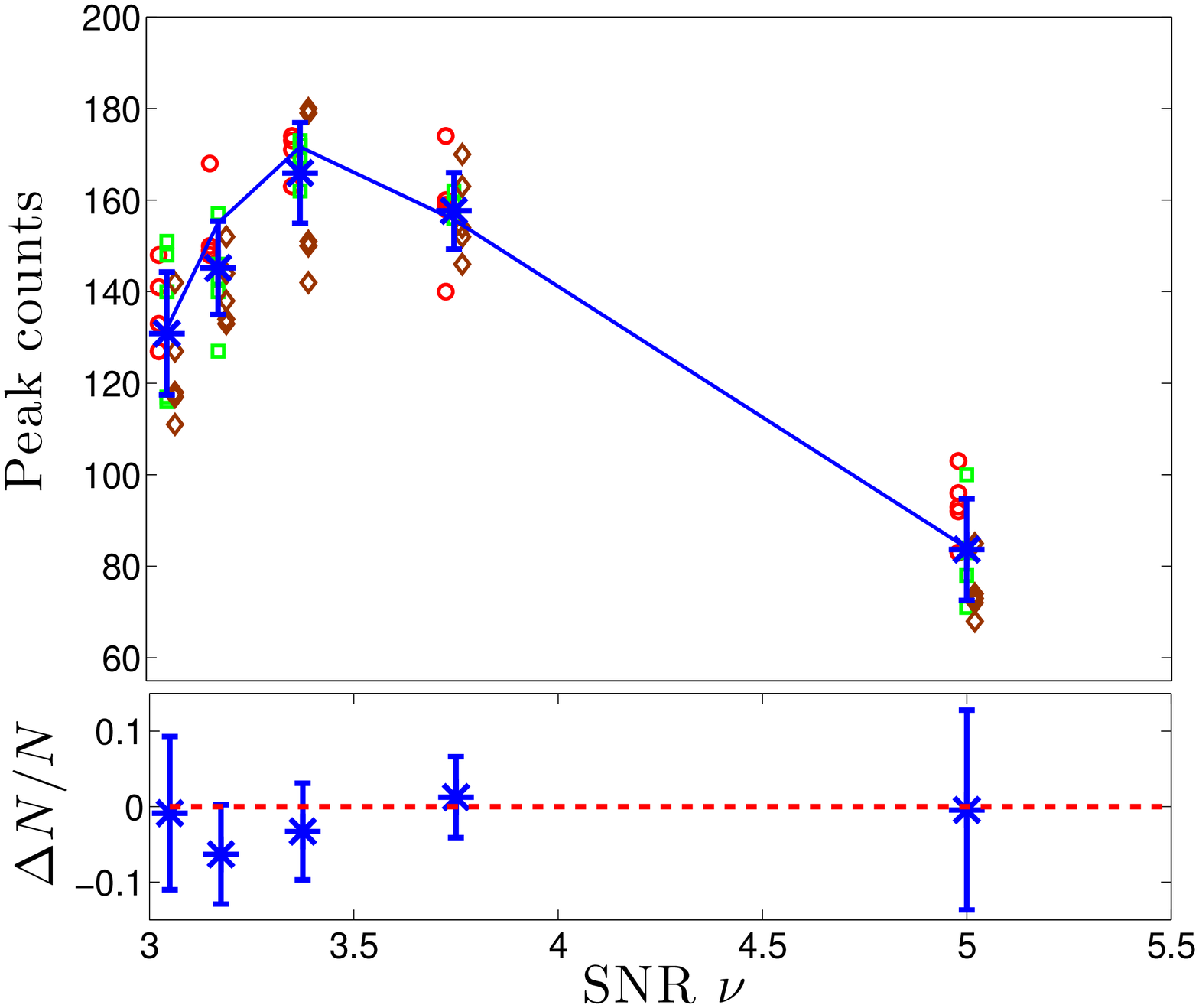}
\caption{CS82 mock simulation results. Upper panels show the peak count distribution in logarithmic scale with equal bins (left),
and the peak count distribution in linear scale with unequal bins (right).
The three sets of symbols with different colours correspond to the $3$ sets of independent mocks. Within the same colour, $5$ data sets 
corresponding to $5$ different noise realizations are shown. The blue `*' and the error bars
are for the average values and the rms over the 15 mocks. The solid line is for our model predictions.
The lower panels show the relative differences between the average values of the $15$ mock sets and our model predictions.
The error bars are for the rms of the relative differences between each mock and the model predictions.
}
\label{fig:mock_CS82}
\end{figure*}

\begin{figure*}
\includegraphics[width=0.48\textwidth]{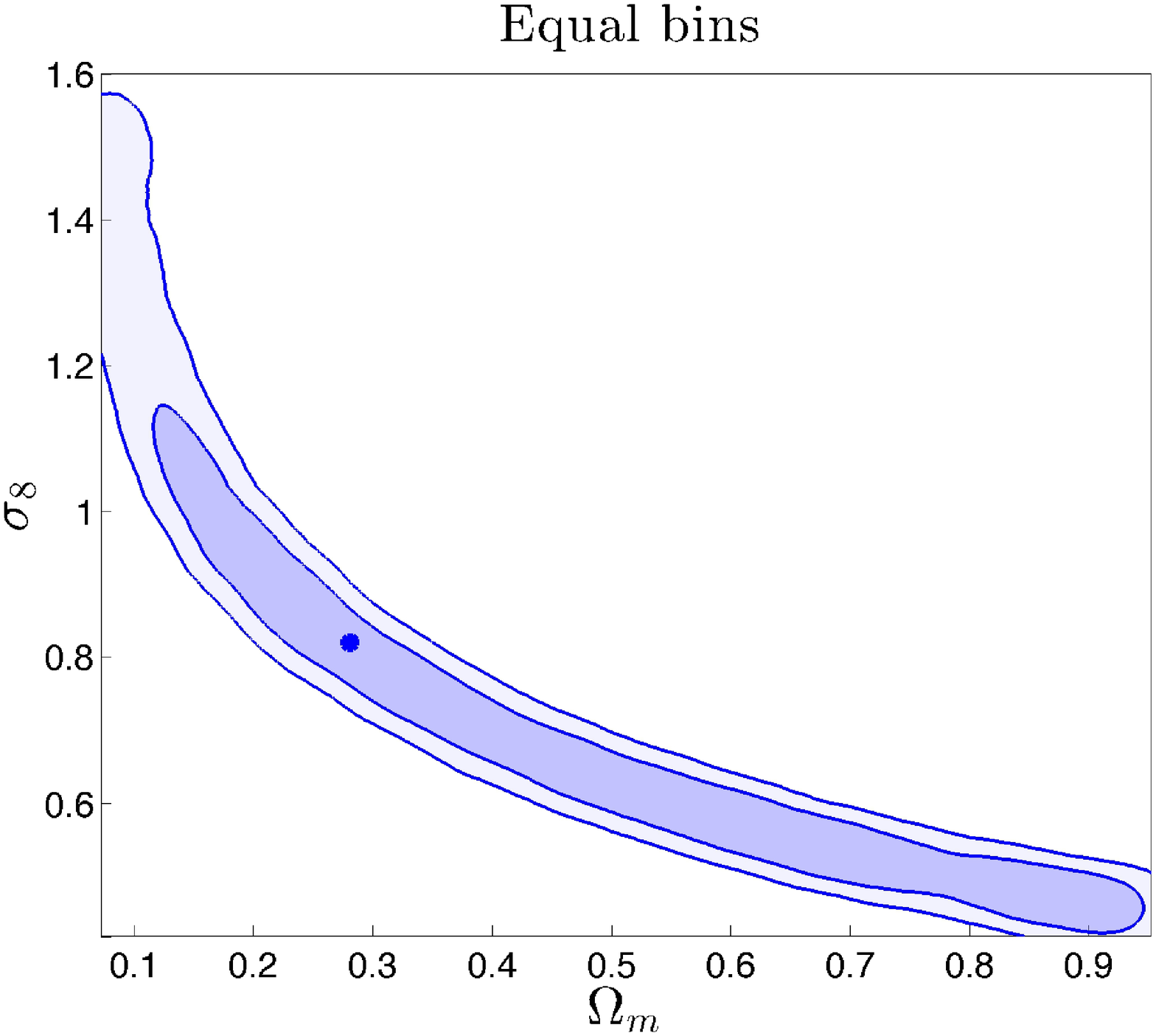}
\includegraphics[width=0.48\textwidth]{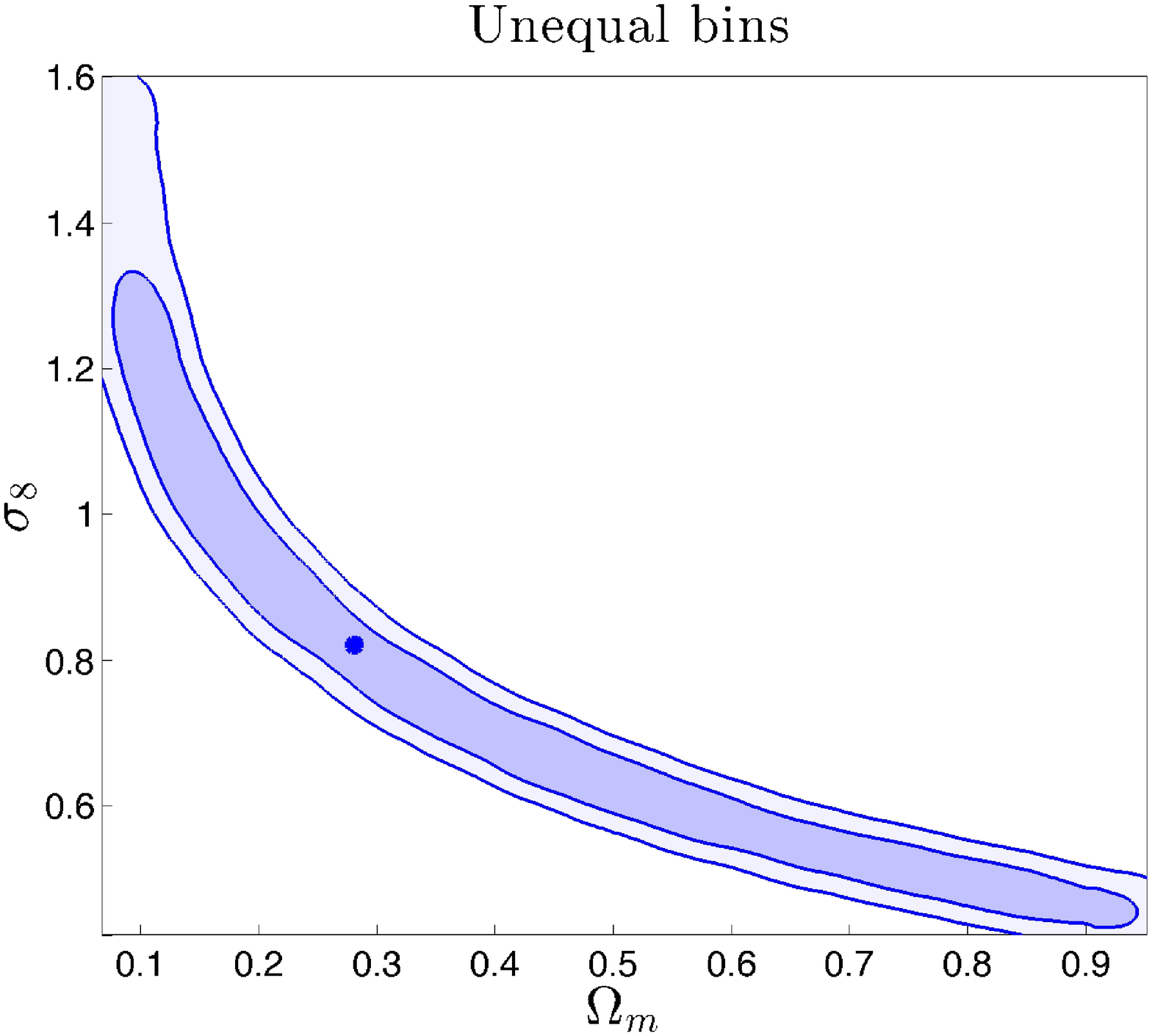}
\caption{Cosmological constraints on ($\Omega_{\rm m}$, $\sigma_8$) derived from the CS82 mock peak counts 
with equal bins (left panel) and with unequal bins (right panel), respectively. The blue dot is for the mock simulation input.}
\label{fig:contour2para_CS82mock}
\end{figure*}

For each mock, the generating procedure is as follows.

(i) With the $3\times 49\deg^2$ lensing maps, we place the tiles of CS82 observed galaxies behind. We note again that 
we do not attempt to mimic the true CS82 long stripe survey geometry here. Therefore we pad the CS82 tiles randomly
over the simulated map area. In each tile, the positions and the amplitudes of ellipticities of the galaxies are preserved, but with
their orientations being randomised. Because there is no exact redshift information for each galaxy, we assign redshifts to the galaxies
following the redshift distribution of Eqn. (\ref{pz}). The galaxy weights and the mask information are also 
preserved in each tile. 

(ii) For a galaxy, its reduced shear $\boldsymbol g$ is calculated by interpolating the signals from the 
pixel positions on simulated maps to the galaxy position. The interpolation is also done in the redshift dimension.  
Regarding the randomised ellipticity obtained in (i) as its intrinsic ellipticity, we then can construct the mock observed 
ellipticity for the galaxy by Eqn.(\ref{eobs}).

(iii) For each tile of the mock data, we perform the convergence reconstruction with the same procedure for the observed data described in \S 3.2
except we do not correct for the multiplicative error in the mock data because our lensing signals are from simulations. 

(iv) We perform the peak identifications and peak number counting in the same way as for the observational data described also in \S 3.2.

In our analyses, noise peaks from shape noise are accounted for. Therefore to obtain a good estimate of the average numbers of peaks,
we randomly rotate the galaxies $5$ times leading to $5$ realizations of the intrinsic ellipticities for source galaxies.  
Therefore, we totally generate $3\times 5$ mocks for CS82.  

\begin{figure*}
\includegraphics[width=0.45\textwidth]{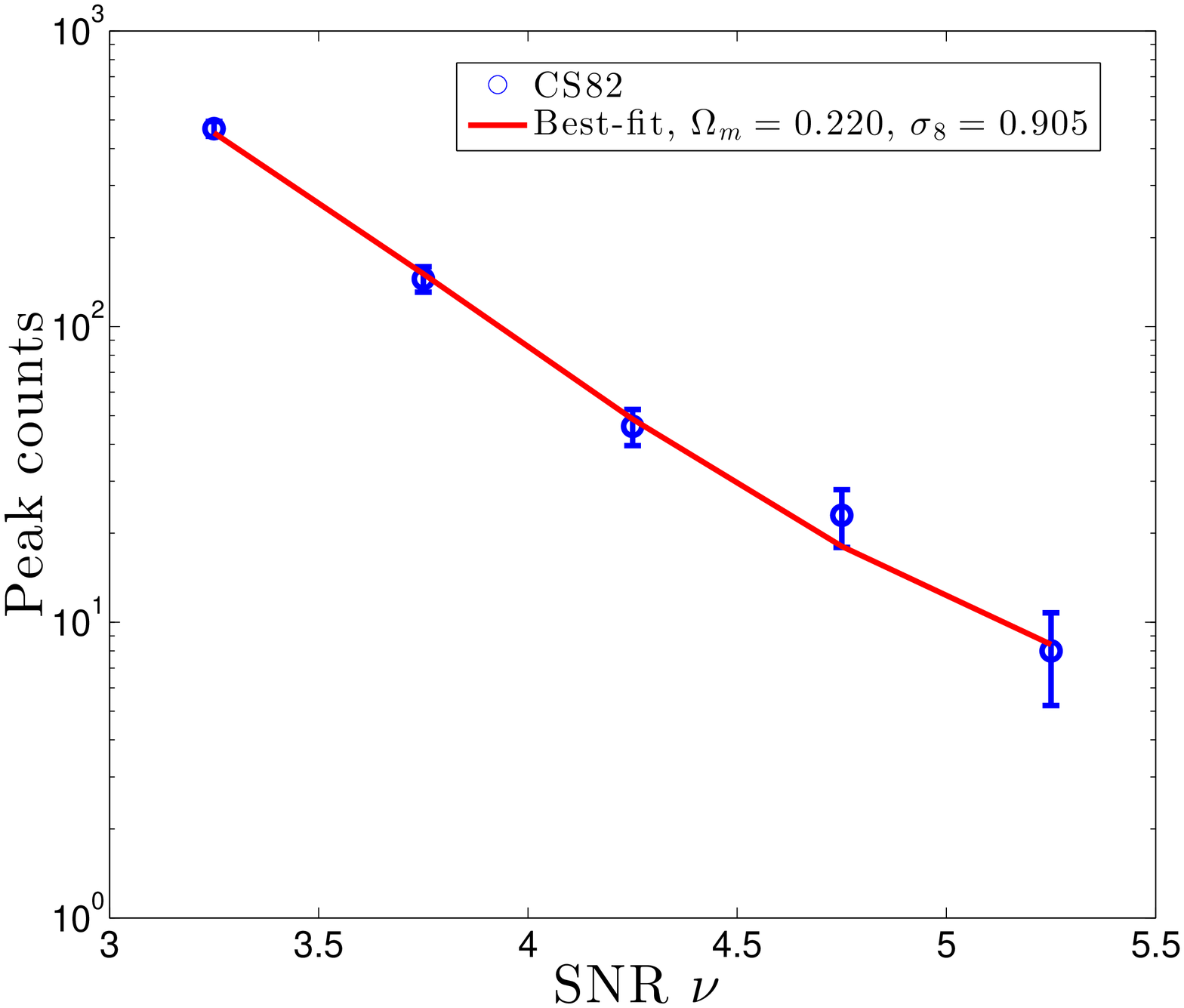}
\includegraphics[width=0.45\textwidth]{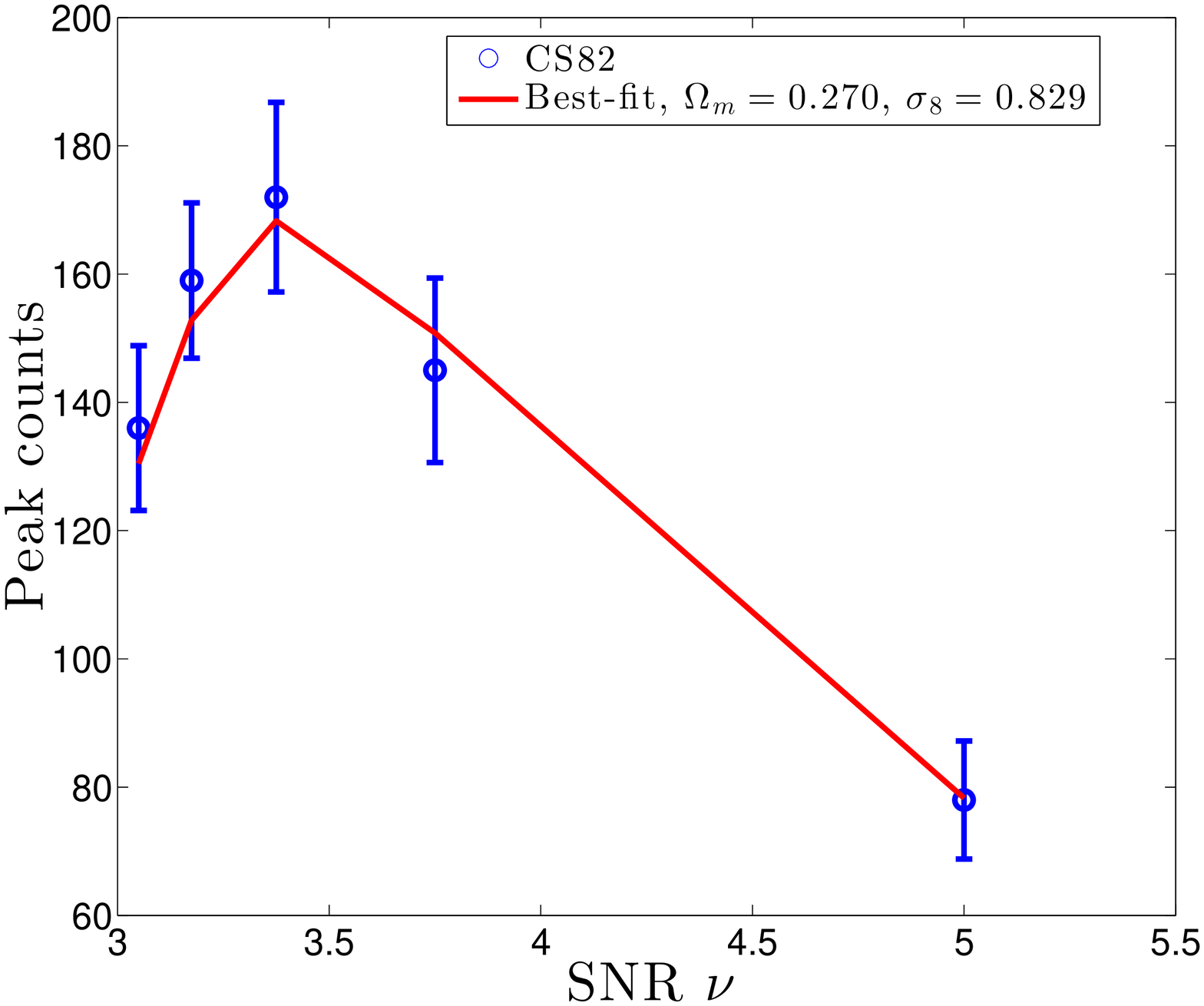}
\caption{CS82 observational results. Left panel: The peak count distribution in logarithmic scale with equal bins. 
Right panel: The peak count distribution in linear scale with unequal bins. 
The corresponding solid line is the theoretical prediction with the best-fit cosmological parameters obtained from MCMC fitting.
The error bars are the square root of the diagonal terms of the covariance matrix.}
\label{fig:peak_CS82}
\end{figure*}

In Figure~\ref{fig:mock_CS82}, we show the peak number distributions for the mock data. The upper left panel 
is for the results (in logarithmic scale) of equal bins with 
the bin width $\Delta\nu=0.5$ in the range of $[3.,5.5]$ and 
the upper right panel (in linear scale) is for the case of unequal bins with $\nu=[3, 3.1],(3.1, 3.25],(3.25, 3.5],(3.5, 4],(4, 6]$, respectively. 
The three sets of symbols with different colours correspond to the mocks from $3$ independent sets of simulated maps. 
The $5$ data points within each colour are the results from different shape noise realizations. The blue '*' and error bars
are for the average values and rms over the $15$ mocks. The solid line is for our model predictions with the 
shape noise level $\sigma_0$ taken to be the average value over all the tiles. The lower panels show the 
corresponding relative differences between the average values of the $15$ mocks and our model predictions.
It is seen that for both binning cases, the averaged mock results agree with our model predictions very well. 
The relative differences are $\le 10\%$, and most often $\le 5\%$.

\begin{table}
\caption{Constraints on $\Omega_{\rm m}$ and $\sigma_8$ from CS82 weak lensing peak abundances. The best-fit values and
the marginalised 1-d mean are shown. The errors are 68\% confidence intervals.}
\label{tab:constraint}
\begin{center}
  \leavevmode
    \begin{tabular}{c c c c c} \hline
                Parameter &  Best fit  &   & 1-d mean &  \\
                          &  (equal bin) & (unequal bin) & (equal bin) & (unequal bin)\\
               \hline
               $\Omega_{\rm m}$  &  $0.22$  &      $0.27$ &       $0.37^{+0.30}_{-0.24}$ & $0.38^{+0.27}_{-0.24}$\\
               $\sigma_8$  &  $0.91$  &      $0.83$ &       $0.83^{+0.28}_{-0.28}$ & $0.81^{+0.26}_{-0.26}$\\
               \hline
     \end{tabular}
    \end{center}
\end{table}

In Figure~\ref{fig:contour2para_CS82mock} we show the derived constraints on $(\Omega_{\rm m}, \sigma_8)$. 
Here we use the results averaged over the $15$ mocks as the `observed' data, and the covariance matrix is estimated by 
constructing bootstrap samples each containing $173$ tiles as the CS82 data from all the tiles in the $15$ mocks. 
The contours are for $1\sigma$ and $2\sigma$ confidence levels. The left and right panels are for the cases of equal bins and unequal bins, respectively.
The blue symbol indicates the underlying parameters of the mocks. We see that noting the strong degeneracy of the two parameters, 
the mock constraints recover the input cosmological parameters excellently. The results from the two binning cases agree well.  

\begin{figure}
\centering
\includegraphics[width=0.95\columnwidth]{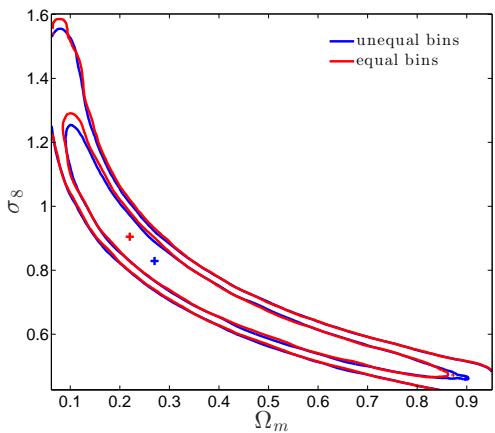}
\caption{Cosmological parameter constraints for ($\Omega_{\rm m}$, $\sigma_8$) derived from CS82 observational peak counts 
with equal bins (red) and with unequal bins (blue), respectively.} 
\label{fig:contour2para_CS82}
\end{figure}

\begin{table}
\caption{Constraints on $\Sigma_8$ and $\alpha$. The errors are 68 per cent confidence intervals.
The 2-pt and 3-pt values are the results in \citet{Fu2014} derived from COSEBis, a second-order E-/B-mode measure and the diagonal third-order aperture mass 
moment, respectively.}
\label{tab:degeneracy}
\begin{center}
  \leavevmode
    \begin{tabular}{c c c c c} \hline
                Parameter         &     equal bin  &    unequal bin  &   2-pt  & 3-pt \\
               \hline
               $\Sigma_8$  &    $0.82\pm0.04$  &      $0.82\pm0.03$ &       $0.79\pm 0.06$ & $0.73^{+0.09}_{-0.19}$\\
               $\alpha$    &    $0.43\pm0.02$  &      $0.42\pm0.02$ &       $0.70\pm 0.02$ & $0.58\pm 0.02$\\
               \hline

     \end{tabular}
    \end{center}
\end{table}

\begin{figure}
\includegraphics[width=\columnwidth]{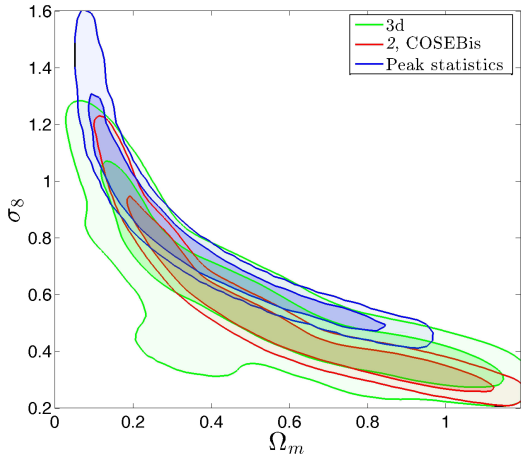}
\caption{The comparison for the constraints on ($\Omega_{\rm m}$, $\sigma_8$) between our peak analyses and the results from 2-pt and 3-pt analyses of \citet{Fu2014}.}.
\label{fig:2pt_compare}
\end{figure}

\begin{figure}
\includegraphics[width=0.95\columnwidth]{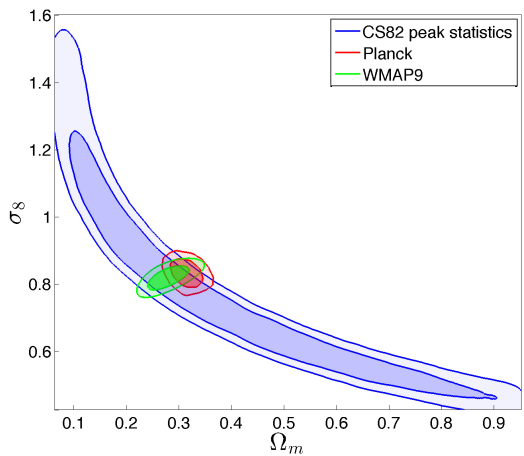}
\caption{Constraints from WMAP9 (green) and Planck (red) overplotted on ours from peak counts with unequal bins.
The contours are for $1\sigma$ and $2\sigma$ confidence levels.} 
\label{fig:contour2para_CS82_combine}
\end{figure}

\subsection{Results from CS82 observational data}

We now proceed to show the results from the CS82 observational data. In Figure~\ref{fig:peak_CS82}, the left and right panels show the results of peak counts
for the equal and unequal bins, respectively. The error bars are from the bootstrap sampling using the CS82
observational tiles. The solid line in each panel is from our theoretical model with the best fit cosmological parameters (see Figure~\ref{fig:contour2para_CS82}). 
In Figure~\ref{fig:contour2para_CS82}, we show the cosmological constraints from the CS82 observed weak lensing peak counts with the red and blue contours
from equal and unequal bins, respectively. The constraints from the two binning cases are consistent with each other very well. 
The best-fit results and the marginalised 1-d mean for $\Omega_{\rm m}$ and $\sigma_8$ are shown in Table~\ref{tab:constraint}.

It is seen that similar to weak lensing correlation analyses \citep[e.g.,][]{Kilbinger2013, Fu2014, Kitching2014}, 
the constraints on the two parameters from weak lensing peak abundances alone 
are strongly degenerate. Considering the relation defined by $\Sigma_8=\sigma_8(\Omega_{\rm m}/0.27)^{\alpha}$ and
using the same estimation method as in \cite{Fu2014}, we obtain $\Sigma_8=0.82\pm 0.04$ and $\alpha=0.43\pm 0.02$ and 
$\Sigma_8=0.82\pm 0.03$ and $\alpha=0.42\pm 0.02$ for the two binning cases, respectively. 
In Table~\ref{tab:degeneracy}, we list the constraints on $\Sigma_8$ and $\alpha$ from our peak analyses and the results from 
2-pt (COSEBis, a second-order E-/B-mode measure) and 3-pt (3d, diagonal third-order aperture-mass moment) analyses 
of \cite{Fu2014} using CFHTLenS data. 
It is noted that for both CS82 and CFHTLenS, the survey areas used in the analyses are $\sim 120\deg^2$. The
mean redshift for CS82 data used in our analyses is $z\sim 0.83$, and the mean redshift for CFHTLenS used in \cite{Fu2014} is
$z\sim 0.75$. It is seen that the $\alpha$ value from our peak analyses is significantly smaller than that from 2-pt and 3-pt analyses.  
This shows a great potential of weak lensing peak analyses in cosmological studies.
In Figure \ref{fig:2pt_compare}, we show the corresponding constraints demonstrating the differences visually.  

We note that the above 2-pt COSEBis and 3-pt 3d analyses from \cite{Fu2014} are in $\Lambda$CDM model with
five free parameters $(\Omega_{\rm m}, \sigma_8, \Omega_{\rm b}, n_{\rm s}, h)$. Their results shown in Table~\ref{tab:degeneracy} 
and Figure \ref{fig:2pt_compare} here are the results marginalised over $(\Omega_{\rm b}, n_{\rm s}, h)$. On the other hand, 
in our peak analyses, we fix the other three parameters and vary only $(\Omega_{\rm m}, \sigma_8)$.
To see if the constraints, particularly the degeneracy direction between $(\Omega_{\rm m}, \sigma_8)$, can be affected
significantly by allowing more free cosmological parameters, we study the dependence of the peak abundances on 
$(\Omega_{\rm m}, \sigma_8, \Omega_{\rm b}, n_{\rm s}, h)$ by calculating the derivatives with respect to these parameters using
the model of F10. It is found that the dependences of peak abundances on $\Omega_{\rm m}$ and $\sigma_8$ are much stronger
than the dependences on the other three parameters. Therefore we do not expect that the inclusion of $(\Omega_{\rm b}, n_{\rm s}, h)$
as free parameters can change our results on the constraints of $(\Omega_{\rm m}, \sigma_8)$ from peak analyses considerably. 
We further perform a-not-so-rigorous Fisher
analysis using the derivatives of the peak abundances with respect to the five parameters and errors corresponding to CS82 data.  
Assuming Gaussian priors for $\Omega_{\rm b}$, $n_{\rm s}$ and $h$ with $\sigma_{\Omega_{\rm b}}=0.05$, $\sigma_{n_{\rm s}}=0.1$ 
and $\sigma_h=0.3$, we find that the constraints on $(\Omega_{\rm m}, \sigma_8)$ marginalised over $(\Omega_{\rm b}, n_{\rm s}, h)$ are
about the same as the results with only $(\Omega_{\rm m}, \sigma_8)$ as free parameters. This shows again that our results 
should not be affected considerably if we include $(\Omega_{\rm b}, n_{\rm s}, h)$ in our peak analyses.

It is noted that the degeneracy of $(\Omega_{\rm m}, \sigma_8)$ from our peak analyses is comparable to the constraints from cluster studies.
From SZ cluster abundance analyses, $\alpha\sim 0.3$ has been obtained \citep[e.g.,][]{SPT2013, ACT2013,Planck2014xx}. 
For X-ray cluster studies, $\alpha\sim 0.5$ \citep[e.g.,][]{Vikhlinin2009, Bohringer2014}. Using SDSS MaxBCG cluster catalogue, 
\cite{Rozo2010} derived $\alpha\sim 0.41$. Our constraint is $\alpha\sim 0.42$, which is in good agreement with the cluster studies 
noting the variations between different analyses largely due to the different observable-mass relation. This is expected
because high weak lensing peaks have close associations with clusters of galaxies along lines of sight.

In the very recent studies of LPH2015, they obtain $\Sigma_8=\sigma_8(\Omega_{\rm m}/0.27)^{0.60}=0.76^{+0.07}_{-0.03}$
using the peak analyses alone combining the results from two smoothing scales. The $\alpha$ value
is larger than ours of $\alpha=0.42$ and $\Sigma_8$ is somewhat smaller. There are a number of differences between their analyses and
ours. They use the peaks spanning a large range of $\kappa$ value from negative to positive.   
In our cosmological studies, we only consider high peaks with $\nu\ge 3$, which corresponds to $\kappa\ge 0.066$ for $\theta_G=1.5\hbox{ arcmin}$
(corresponding to $\sim 1\hbox{ arcmin}$ in LPH2015 because of the different definition of $\theta_G$). 
We expect that our results should resemble more those of cluster studies as explained in the previous paragraph. For low peaks, 
besides the impact of noise, they are related to the projection effects of large-scale structures, which 
might be largely described by the underlying power spectrum. 
However, in LPH2015, they find that although weak lensing peaks show
stronger covariance with the power spectrum than cluster counts, the overall covariance between peaks and the power spectrum is
rather weak. Therefore even low peaks and the power spectrum should contain non-overlapping cosmological information.
The error covariance matrix estimation is also different in the two studies. We use 
the bootstrap approach by using the data themselves. In LPH2015, they use a fiducial simulation to calculate the covariance 
by randomly rotating and shifting the simulation box of size $240h^{-1}{\rm Mpc}$ in ray tracing.  To fully explore the cause of the differences from different analyses can be a worthwhile 
task in the future.

In Figure~\ref{fig:contour2para_CS82_combine}, we show the constraints in comparison with the results from WMAP9 \citep{Hinshaw2013} (green) and Planck 
\citep{Planck2014} (red). It can be seen that our constraints are in good agreement with both.

\begin{figure}
\includegraphics[width=0.95\columnwidth]{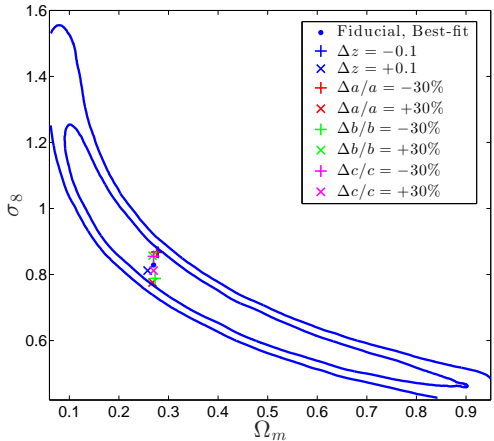}
\caption{The impact of the uncertainties in the galaxy redshift distribution on cosmological parameter constraints derived from CS82 weak lensing
peak abundances.}
\label{fig:zerrortest}
\end{figure}

The above analyses adopt the fiducial redshift distribution given in Eqn.~(\ref{pz}) with $a=0.531$, $b=7.810$ and $c=0.517$ for CS82 galaxies.
Derived by matching to COSMOS galaxies, this redshift distribution can have significant uncertainties. To understand the impact of the uncertainties 
on our peak analyses, we follow \cite{Hand2013} to vary the parameters in the redshift distribution and analyse how the peak abundances change
using the theoretical model of F10. Considering peaks in the range of $\nu=[3,6]$, we find that by shifting the peak position of the redshift distribution 
by $\Delta z=\pm 0.1$, the peak abundances change by $\sim \pm 3\%$ to $\sim \pm 20\%$ from low to high peaks. 
Varying the $b$ parameter by $+30\% (-30\%)$ leads to $\sim -2\% (+5\%)$ to $\sim -10\% (+20\%)$ changes in the peak abundances. 
A $\pm 30\%$ change in $c$ parameter causes $\le 5\%$ changes in peak abundances. The largest impact is from the uncertainties in $a$ parameter. 
Changing $a$ by $+30\% (-30\%)$ leads to $\sim +8\% (-5\%)$ to $\sim +30\% (-20\%)$ variations in the peak abundances from low to high peaks.
With these varied redshift distributions, we estimate the best fit $\sigma_8$ from observed CS82 peak abundances (unequal bins)
by keeping $\Omega_{\rm m}$ to be approximately the best fit value from our fiducial analyses. The results are shown in Figure~\ref{fig:zerrortest}. 
It is seen that for all the above changes, the best fit $\sigma_8$ are within the $1\sigma$ range of our fiducial constraints. 
Therefore we do not expect a considerable impact on our results from the uncertainties in the galaxy redshift distribution.   
On the other hand, for future surveys with dramatically improved statistics, the uncertainties in the redshift distribution can be 
an important source of systematic errors for weak lensing peak statistics, and the required accuracy for redshift measurements
needs to be carefully studied.   

\begin{figure*}
\centering
\includegraphics[width=0.45\textwidth]{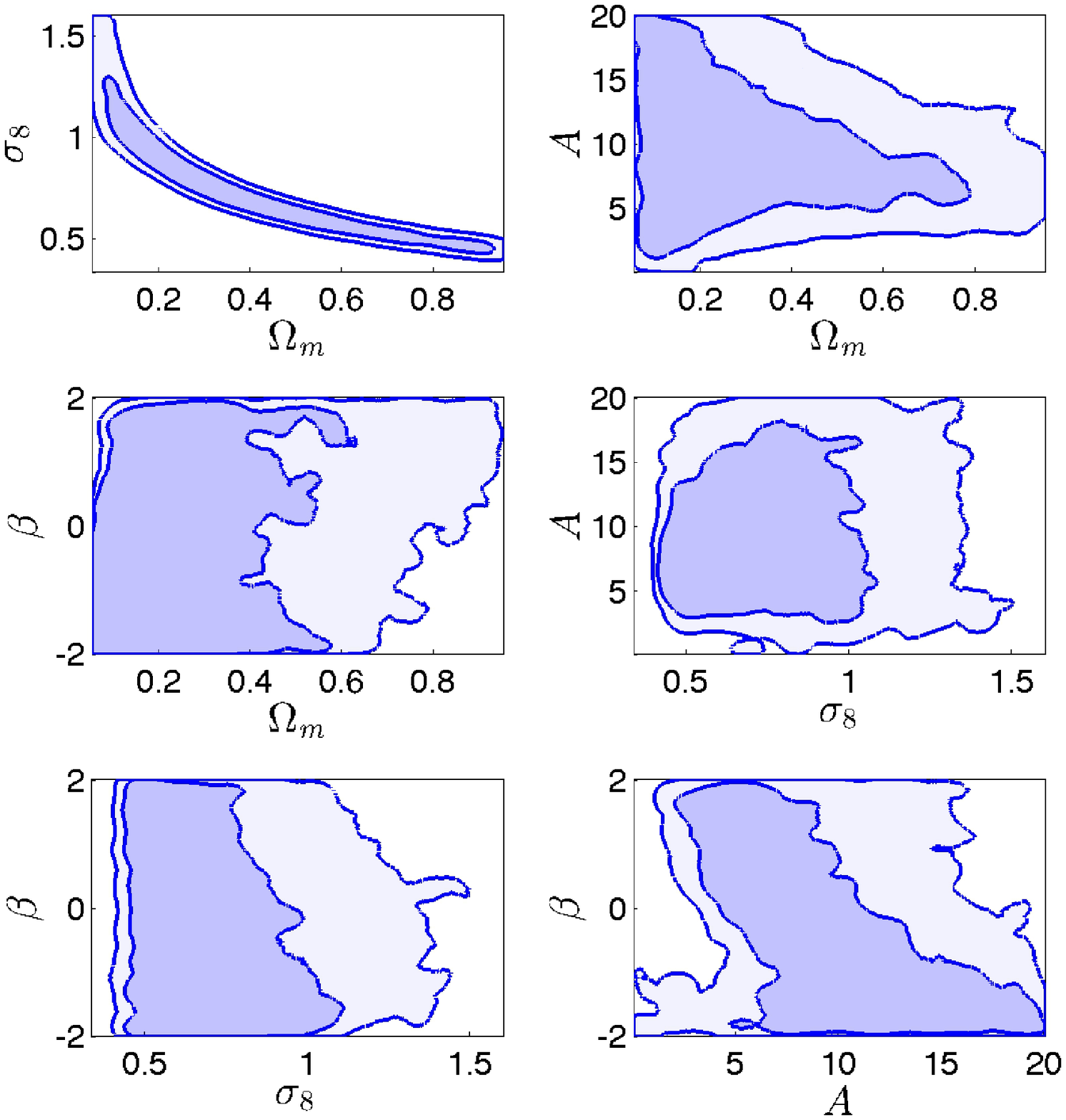}
\includegraphics[width=0.45\textwidth]{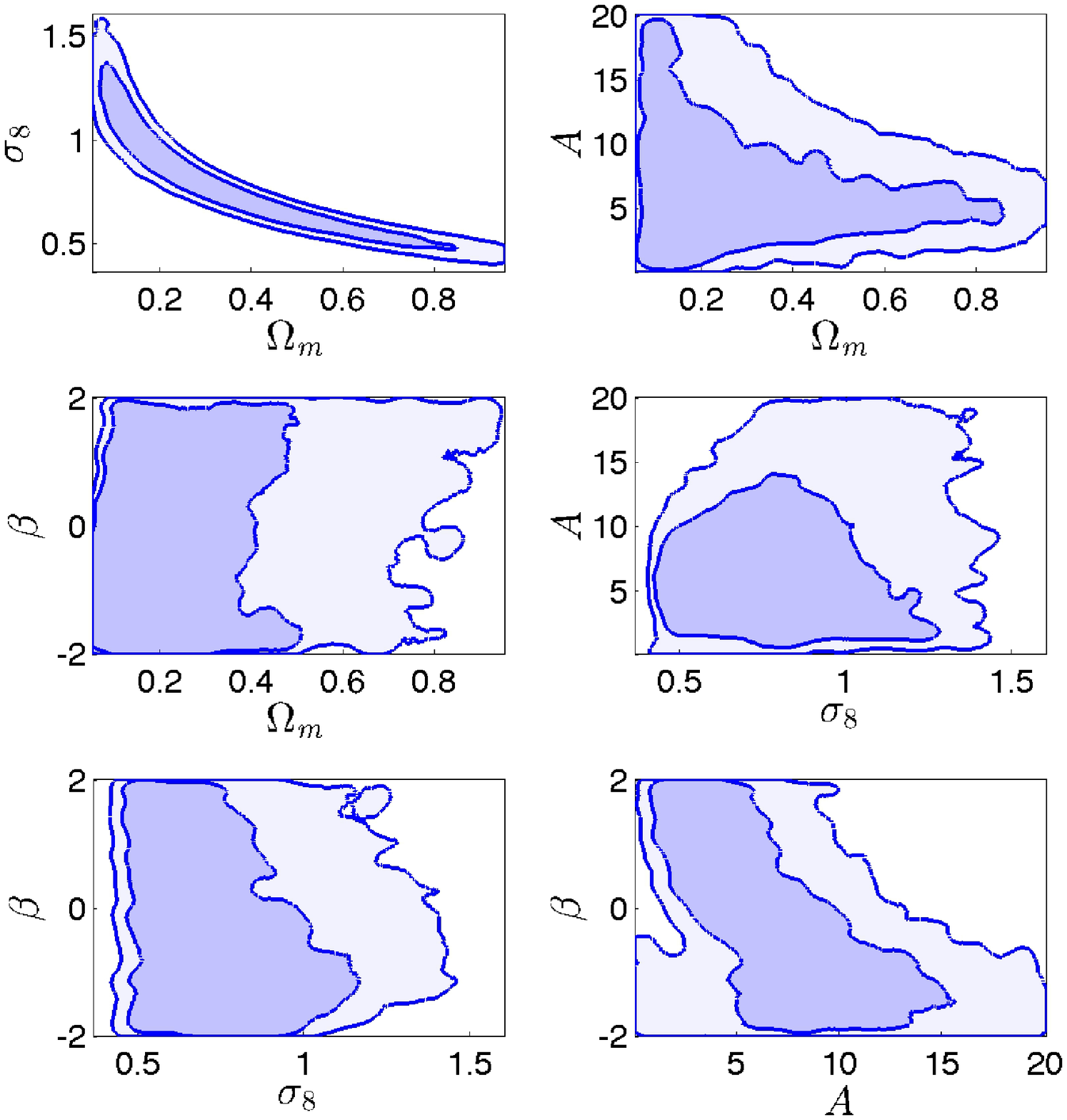}
\caption{The 4-parameter constraints from the mocked (left) and observed (right) unequal binned weak lensing peak counts.
The contours are for $1\sigma$ and $2\sigma$ confidence levels}
\label{fig:contour4para_CS82}
\end{figure*}

\begin{table}
\caption{Marginalised 1-d mean from the 4-parameter fitting using unequally binned peak counts. 
The errors are 68\% confidence intervals.}
\label{tab:4paraconstraint}
\begin{center}
  \leavevmode
    \begin{tabular}{c c c } \hline
                Parameter &  Mock  & CS82 observation   \\
               \hline
               $\Omega_{\rm m}$  &  $0.38^{+0.30}_{-0.26}$ & $0.34^{+0.30}_{-0.23}$\\
               $\sigma_8$   &       $0.80^{+0.28}_{-0.27}$ & $0.84^{+0.28}_{-0.27}$\\
               $A$          &  $9.7^{+5.4}_{-4.8}$ & $7.5^{+5.2}_{-4.6}$ \\
               $\beta$      & $-0.15^{+1.41}_{-1.45}$ & $-0.15^{+1.37}_{-1.31}$ \\ 
               \hline

     \end{tabular}
    \end{center}
\end{table}

\subsection{Other constraints}

Besides cosmological parameter constraints, weak lensing peak statistics may also possibly provide constraints on the 
density profile of dark matter halos because massive dark matter halos are the sources of true weak lensing peaks with 
high signal-to-noise ratios \citep[e.g.,][]{Yang2013}. In our theoretical model of F10, the dependence of the peak abundances on the density profile 
of dark matter halos is explicit. This allows us to perform constraints on the structural parameters of dark matter halos 
simultaneously with cosmological parameters. Similar ideas are also recently proposed in other studies \citep[e.g.,][]{Mainini2014, Cardone2015}. 
To show this feasibility, we assume a power-law mass-concentration relation for NFW halos in our model calculations with \citep[e.g.,][]{Duffy2008}

\begin{equation}
c_{\rm vir}=\frac{A}{(1+z)^{0.7}}\left(\frac{M_{\rm vir}}{10^{14}\mathrm{h^{-1}M_\odot}}\right)^{\beta},
\label{cm}
\end{equation}
where $A$ and $\beta$ are regarded as free parameters. The redshift dependence $(1+z)^{0.7}$ is taken to be 
consistent with recent simulation results \citep[e.g.,][]{Bhattacharya2013}. We then perform 4-parameter fitting 
from the observed weak lensing peak counts. The flat priors are $[0, 20]$ for $A$, $[-2, 2]$ for $\beta$,
$[0.05, 0.95]$ for $\Omega_{\rm m}$, and $[0.2, 1.6]$ for $\sigma_8$. 
The results are shown in Figure~\ref{fig:contour4para_CS82} where the unequal binned peak counts are used. 
The left panels are for the results from mock data described in \S 4.2, and the right 
panels are from CS82 observational data. We see that although the obtained constraints are mainly for $(\Omega_{\rm m}, \sigma_8)$, 
the constraints on the plane of $(A,\beta)$ are apparent even for the current generation of weak lensing surveys. 
The marginalised 1-d mean for both mock and the CS82 observational analyses are shown in Table~\ref{tab:4paraconstraint}. 
The results for $A$ and $\beta$ are in broad agreement with the results from simulated halos although the error ranges are large. 
We also note that in this 4-parameter fitting, the constraint contours on $(\Omega_{\rm m},\sigma_8)$ plane 
are enlarged somewhat in comparison with the results of the 2-parameter fitting case, showing the influence of the uncertainties 
of the halo structural parameters on the cosmological constraints.  

For future surveys with much larger survey areas and the improved depth, statistically we expect significant enhancements
of weak lensing peak analyses, which in turn will provide us valuable cosmological information complementary
to cosmic shear correlation analyses.  


\section{Summary and discussions}
\label{summary}

With CS82 weak lensing observations, we study the weak lensing peak abundances for $\nu$ in the range 
$[3,6]$, and derive the first cosmological constraints from peak analyses. We summarise the results
as follows.

(i) For flat $\Lambda$CDM, the cosmological constraints on $(\Omega_{\rm m}, \sigma_8)$ from peak analyses are fully consistent with the constraints obtained from
cosmic shear correlation studies. On the other hand, the degeneracy direction of the two parameters is flatter than 
those from the correlation analyses. Quantitatively, with $\Sigma_8=\sigma_8(\Omega_{\rm m}/0.27)^{\alpha}$, we obtain
$\alpha=0.42\pm 0.02$ in comparison with $\alpha=0.70\pm 0.02$ from COSEBis 2-pt correlations
and $\alpha=0.58\pm 0.02$ from diagonal three-order aperture-mass correlation studies \citep{Fu2014}. 
This shows a promising potential of weak lensing peak analyses complementary to correlation studies. It is noted that to 
explore the improvements on the cosmological parameter constraints from the combined analyses, the full covariance between  
the peak abundances and the correlation functions should be investigated carefully. 

(ii) Our derived cosmological constraints from peak analyses are also consistent with both WMAP9 and Planck results.

(iii) We perform constraints on $(A, \beta)$, the power-law form of the mass-concentration relation of dark matter halos, simultaneously 
with the cosmological parameters $(\Omega_{\rm m}, \sigma_8)$. For the CS82 survey with relatively large statistical errors, the current
constraints are mainly on $(\Omega_{\rm m}, \sigma_8)$. However, the constraints on $(A, \beta)$ are already apparent. This 
shows the capability to constrain the structural parameters of massive structures together with cosmological parameters
from weak lensing peak statistics. With much improved data from future surveys, performing simultaneous constraints on the
structural and cosmological parameters can potentially allow us to extract important astrophysical effects on the structural evolution of 
the mass distribution of halos \citep[e.g.,][]{Yang2013}. 
Meanwhile, it can also avoid the possible bias on cosmological parameter constraints resulting from 
the pre-assumption about the halo structures in predicting the cosmological dependence of weak lensing peak abundances. 

In this paper, we adopt the theoretical model of F10 for predicting the peak abundances. The model takes into account the effects of shape noise
in the calculation, and has been tested extensively with ray-tracing simulations in our previous studies (F10, LWPF2014). 
For CS82, the shape noise is the dominant source of error. The applicability of the model
is further shown with our mock analyses presented in \S 4.2 here. Comparing to the approach fully relying on large simulations,   
theoretical modelling can help us understand better different effects, and can allow us to explore cosmological- and 
dark matter halo density profile-parameter space efficiently.  

On the other hand, for future surveys with much reduced statistical errors, our theoretical model needs to be developed to 
include the effects neglected in the current treatment, such as the projection effects of large-scale structures, the 
complex mass distribution of halos, etc.. Unlike the shape noise, such effects themselves also contain cosmological information.
With the help of simulations, we are currently working toward improving our model for future cosmological applications.

Weak lensing effects are unique in probing the dark side of the Universe. Current generation of completed surveys, such as CS82 and CFHTLenS, 
have served as important demonstrations to show the feasibility of weak lensing cosmological studies. Ongoing surveys, such as DES \citep{DES2005}, 
HSC \citep{HSC2009}, and KiDS \citep{KIDS2013}, will expand the survey area to a few thousands square degrees. Future ones, such as Euclid \citep{Euclid2012}
and LSST \citep{LSST2009}, will target at nearly half of the sky of about $20,000 \deg^2$. The statistical capability of weak lensing studies 
will increase tremendously. To fully realise the power, however, different systematics, both observational and theoretical ones, 
need to be understood thoroughly. For this, the current surveys also play important roles in revealing different obstacles that 
need to be overcome and further paving the road to the future. 


\section*{Acknowledgements}

Based on observations obtained with MegaPrime/MegaCam, a joint project of CFHT and CEA/DAPNIA,
at the CFHT, which is operated by the National Research Council (NRC) of Canada, the Institut National des Science de l'Univers of
the Centre National de la Recherche Scientifique (CNRS) of France and the University of Hawaii. The Brazilian partnership on CFHT
is managed by the Laborat\'orio Nacional de Astrof\'isica (LNA). This work made use of the CHE cluster, managed and funded by ICRA/CBPF/MCTI,
with financial support from FINEP and FAPERJ. We thank the support of the Laboratrio Interinstitucional de e-Astronomia (LIneA).
We thank the CFHTLenS team for their pipeline development and verification upon which much of this surveys pipeline was built.

We thank the referee for the constructive comments and suggestions.
We thank the useful discussions with Z. Haiman and J. Liu. XKL, CZP and ZHF acknowledge the support from NSFC of China under the grants
11333001, 11173001, and 11033005, and from the Strategic Priority Research Program ``The Emergence of Cosmological Structures''
of the Chinese Academy of Sciences, Grant No. XDB09000000.
HYS acknowledges the support by a Marie Curie International Incoming Fellowship within the $7^{th}$ European Community 
Framework Programme, and NSFC of China under grants 11103011.
LPF acknowledges the support from NSFC grants 11103012 and 11333001, and Shanghai Research grant 13JC1404400 of STCSM.
LR acknowledges the support from NSFC grant 11303033 and the support from Youth Innovation Promotion Association of CAS.
AL acknowledges the support by World Premier International Research Centre Initiative (WPI Initiative), MEXT, Japan.
JPK acknowledges support from the ERC advanced grant LIDA and from CNRS.
BM acknowledges financial support from the CAPES Foundation grant 12174-13-0.
MM is partially supported by CNPq (grant 486586/2013-8) and FAPERJ (grant E-26/110.516/2-2012).

The MCMC calculations are partly done on the Laohu supercomputer at the Centre of Information and Computing at 
National Astronomical Observatories, Chinese Academy of Sciences, funded by Ministry of Finance of Peoples Republic of China 
under grant ZDYZ2008-2. The N-body simulations are performed on the Shuguang cluster at Shanghai Normal University, Shanghai, China.


\bibliography{ms}

\section*{Appendix}

In this appendix, we describe our programming structure for computing theoretically the peak abundances using the model of F10.
For clarity, we copy the relevant equations in \S 3.3 here. For a signal-to-noise ratio $\nu$, 
the surface number density of peaks is given by   
\begin{equation}
n_{\mathrm{peak}}(\nu)d\nu=n_{\mathrm{peak}}^c(\nu)d\nu+n_{\mathrm{peak}}^n(\nu)d\nu.
\label{app_npeaktwoterm}
\end{equation}
The part for pure noise peaks $n_{\mathrm{peak}}^n(\nu)$ can be calculated easily if the regions occupied by halos 
are known. Therefore the most computationally heavy part is $n_{\mathrm{peak}}^c(\nu)$ for peaks in halo regions. 
In F10, it can be calculated by
\begin{equation}
 n_{\mathrm{peak}}^c(\nu)=\int{dz\frac{dV(z)}{dzd\Omega}}\int_{\rm M_{\rm lim}}{dMn(M,z)f_p(\nu,M,z)},
\label{app_npeakc}
\end{equation}  
and 
\begin{equation}
f_p(\nu,M,z)=\int_{0}^{\theta_{\mathrm{vir}}}d\theta\hbox{ } (2\pi \theta)\hbox{ } \hat {n}^c_{\mathrm{peak}}(\nu,\theta,M,z),
\label{app_fnumz}
\end{equation}
where
\begin{eqnarray}
&&\hat n^c_{\mathrm{peak}}(\nu,\theta, M, z)=\exp \bigg [-\frac{(K^1)^2+(K^2)^2}{\sigma_1^2}\bigg ]\nonumber \\
&&\times \bigg [ \frac{1}{2\pi\theta_*^2}\frac{1}{(2\pi)^{1/2}}\bigg ]
\exp\bigg [-\frac{1}{2}\bigg ( \nu-\frac{K}{\sigma_0}\bigg )^2\bigg ] \nonumber \\
&&\times \int_0^{\infty} dx_N\bigg \{ \frac{1}{ [2\pi(1-\gamma_N^2)]^{1/2}}\nonumber \\
&&\times \exp\bigg [-\frac{ [{x_N+(K^{11}+K^{22})/ \sigma_2
-\gamma_N(\nu_0-K/\sigma_0)}]^2}{ 2(1-\gamma_N^2)}\bigg ] \nonumber \\
&& \times  F(x_N)\bigg \}.
\label{app_nchat}
\end{eqnarray}
It is noted that given a pair of $(M, z)$ for a halo, the function $\hat {n}^c_{\mathrm{peak}}(\nu,\theta,M,z)$ 
in Eqn.~(\ref{app_nchat}) can be computed independently for different $\theta$ and $\nu$. We therefore employ GPU
for this part of calculations, which improves our computational efficiency enormously. 

\begin{figure}
\centering
\includegraphics[width=0.95\columnwidth]{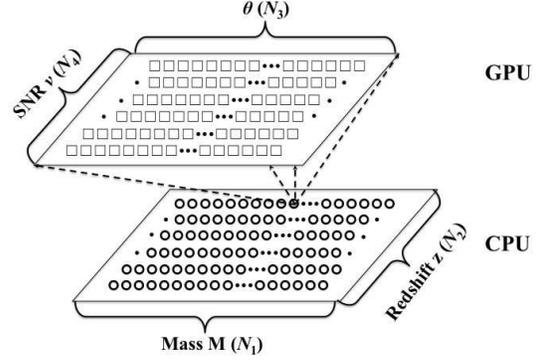}
\caption{Sketch for our code calculations.}
\label{fig:gpu}
\end{figure}

We illustrate our programming structures in Figure~\ref{fig:gpu}. The specifics are as follows.

\begin{enumerate}
\item We first divide the halo mass range and the redshift range into $N_1\times N_2$ grid points as shown in the bottom layer of
Figure~\ref{fig:gpu}. Memories are allocated for mass $M$ ($N_1$ sample points) and redshift $z$ ($N_2$ sample points) matrix on host memory (CPU memory). 
We then calculate the corresponding virial radius $r_{\rm{vir}}(M,z)$ on CPU according to Eqn.(\ref{virialr}).

\item Then the results of $r_{\rm{vir}}$ are copied to global memory (GPU memory). We allocate global memory for different 
$\theta$ ($N_3$ sample points) and SNR $\nu$ ($N_4$ sample points) as shown in the top layer of Figure~\ref{fig:gpu}.

\item We take the advantage of GPU shared memory to evaluate the smoothed convergence field $K$ and its first and second derivatives for an NFW halo
using Chebyshev interpolation algorithm for integration \citep{Fox1968}. The shared memory is highly efficient and suitable to use in this situation.

\item Launch the main kernel to evaluate $f_{\rm p}(\nu, M ,z)$ according to Eqn. (\ref{app_fnumz}) and Eqn.(\ref{app_nchat}). 
This involves massive amount of calculations because of the multi-dimensional integrations and is benefited greatly by using the GPU feature .

\item At last, we copy the results of $f_{\rm p}(\nu,M, z)$ from global memory to host memory, and calculate the final result of $n_{\mathrm{peak}}^c(\nu)$
using the 11-th order Simpson method on CPU \citep{Press2007}. 

\end{enumerate}

The use of GPU improves the model calculation efficiency significantly. For example, our own workstation
consists of 2 Xeon E5-2697v2 CPUs and 4 Nvidia GTX Titan GPU. Each CPU contains 12 cores while each GPU contains 2880 cores.
In this case, we achieved a calculation speed which is more than 20 times faster by using GPU than that by using CPU alone.

\label{lastpage}
\end{document}